\documentclass[11pt,a4paper]{article}
\usepackage{jheppub}
\usepackage{dsfont}
\usepackage{graphicx}
\usepackage{float}
\usepackage[small]{caption}
\renewcommand\Re{\operatorname{Re}}
\renewcommand\Im{\operatorname{Im}}

\newcommand{\mathsym}[1]{{}}

\title{Degenerate Rotating Black Holes, Chiral CFTs and Fermi Surfaces I -
         Analytic Results for Quasinormal Modes }

\author[a]{Micha Berkooz}
\author[a]{Anna Frishman}
\author[a]{Amir Zait}

\emailAdd{micha.berkooz@weizmann.ac.il}
\emailAdd{anna.frishman@weizmann.ac.il}
\emailAdd{amir.zait@weizmann.ac.il}


\affiliation[a]{
$^1$Department of Particle Physics and Astrophysics,
The Weizmann Institute of Science,
Rehovot 76100, Israel
}

\abstract{
In this work we discuss charged rotating black holes in $AdS_5 \times S^5$ that degenerate to extremal black holes with zero entropy.
These black holes have  scaling properties between charge and angular momentum similar to those of Fermi surface operators in a subsector of $\mathcal{N}=4$ SYM.  We add a massless uncharged scalar to the five dimensional supergravity theory, such that it still forms a consistent truncation of the type IIB ten dimensional supergravity and analyze its quasinormal modes.
Separating the equation of motion to a radial and angular part, we proceed to solve the radial equation using a matched asymptotic expansion method applied to Heun's equation with two nearby singularities. We use the continued fraction method for the angular Heun equation and obtain numerical results for the quasinormal modes. In the case of the near-supersymmetric black hole we present some analytic results for the decay rates of the scalar perturbations. The spectrum of quasinormal modes obtained is similar to that of a chiral $1+1$ CFT, which is consistent with the conjectured field-theoretic dual. In addition, some of the modes can be found analytically.}

\begin{document}
\begin{flushright}
WIS/09/12-JUNE-DPPA\\
\end{flushright}

\maketitle

\section{Introduction}
The AdS/CFT correspondence, first conjectured by Maldacena in his seminal work in 1997 \cite{Maldacena1998},
provides a framework allowing us to probe the quantum nature of gravity on Anti-de Sitter (AdS) spacetimes
via familiar quantum field theory (QFT) tools, and vice-versa.

Black holes, being a striking feature of gravitational theories, have played a particularly important role in this duality and its implications. The reason is that black holes, which have finite entropy, count the generic states in the theory with some fixed charges, whether in AdS or in flat space. Indeed, their study has shed light on the fundamental degrees of freedom of some gravitational background (in near horizon limits), and has the potential to provide more information on yet other backgrounds. If the black hole is supersymmetric, then under favorable circumstances one can more easily compare weak and strong coupling results. In cases without supersymmetry, it is difficult to get an exact matching of weak and strong coupling computations, although in some case one can understand some general scaling properties.

The most intensively studied case of the AdS/CFT correspondence is the duality between type $IIB$ String theory on $AdS_5 \times S^5$ and
$\mathcal{N} = 4\ SU(N)$ Super Yang-Mills (SYM) gauge theory. It will also be the focus of the current paper. More precisely, we concentrate on rapidly rotating black holes in the $U(1)^3$ SUGRA consistent truncation on $AdS_5$. Currently, the most generic black holes known in this theory have equal charges under two of the $U(1)$s, with all the other charges arbitrary \cite{Mei2007} (relying on earlier work in \cite{Cvetic:2004ny,Chong2005,Chong2005b,Chong2007,Kunduri2006}). We will be interested in the dynamical (in)stability of zero entropy degenerations of such black holes. Our results therefore lie at the interface of several questions:

\begin{enumerate}
\item {\it Degeneration of Black holes into singularities}
One of the successes of string theory is to resolve a large class of singularities which are ill defined in gravity. The simplest of these are of course the orbifold theories (for example non-SUSY singularities which are not accessible geometrically, cf. \cite{Adams:2001sv}). Still, there are more singularities which are not understood compared to those that are.

The weak cosmic censorship hypothesis tells us it is difficult to create singularities dynamically, in the sense that generic initial conditions will create at best a singularity cloaked by an horizon. One can turn this the other way around and study backgrounds in which there is a parametrically tiny horizon around a singularity. Such scenarios will hopefully capture the dynamics of the singularity while regulating at least some of the problems associated with studying singular backgrounds.

The black holes we will discuss are of this nature. More precisely one can distinguish two cases - the first, which we will only touch upon briefly, is one where there is a finite entropy horizon and a singularity sits behind the horizon and approaches the horizon at some critical value of a parameter.

The other, which is our main emphasis, is the case where the black hole entropy goes to zero. In this limit the volume factor of the horizon shrinks to zero size uniformly along the horizon, except at a lower dimensional locus where a singularity develops.  Our paper deals with some general properties of quasi- normal modes in such a case.

\item {\it Quasinormal modes of fast rotating black holes}

Probing the linear (in)stability of a black hole background
requires one to compute the allowed frequencies of the modes of perturbing fields.
These frequencies are usually complex, as black holes are dissipative systems,
and are thus called Quasi-Normal Modes (QNMs for short).
Instability is observed when the imaginary part causes the amplitude to grow exponentially,
breaking the linear approximation. Such instabilities are useful when attempting
to understand the phase structure of the black holes, and via the duality,
the phase structure of the field theory at strong coupling.

Computing these quasinormal modes analytically for AdS black holes is a difficult problem.
It has been solved only for black holes which are small compared to the AdS radius,
and are furthermore slowly rotating, cf. ref. \cite{Cardoso2004}.
The black holes discussed in this paper are fast-rotating near-extremal AdS black holes that are comparable in size to the AdS radius.
Perturbations on such backgrounds are hard to study, even numerically \cite{Uchikata2009}.
Here we compute numerically the quasinormal modes of a massless scalar in the extremal limit for
fast rotating black holes. For the near-SUSY black holes, we are able to compute analytically the lowest QNM,
which is the one with slowest decay rate.

The AdS/CFT correspondence relates the QNMs
to the poles of the retarded Green's function in the dual CFT \cite{Son:2002sd}.
Through this relation, they can be used to study decay rates of excitations above a known ground state.
One can therefore hope that having analytic expressions for QNMs will allow for
precise matching to computations carried out on the field theory side.

\item {\it Fermi surfaces of fundamental fermions with the AdS/CFT correspondence}

Fermi surfaces, a robust feature of theories with
fermionic degrees of freedom, are conjectured to have extremal black holes
as their gravity duals. These constructions are rather constrained though. Fermi surfaces are expected to have zero entropy - this is true at least for the Fermi surfaces encountered in most CM systems (with the caveat that these are not large N systems in contrast to the setting of the AdS/CFT correspondence).  Thus, extremal black holes, when taken to the limit in which they degenerate to have zero entropy, would make better candidates for Fermi surfaces.

In this paper we propose a duality between known extremal black hole solutions
in gauged SUGRA and Fermi surface operators living in a sector of the full $\mathcal{N} = 4$ theory.
Several works have suggested using Fermi surface-like
operators as components of known $1/16$ BPS charged and
rotating black holes \cite{Berkooz2007,Berkooz2008}.
It has been shown that while some of the properties of the Fermi surfaces are reproduced by the black holes, others are not. Specifically, these have $N^2$ worth of entropy, an indication that they are mixed states of many operators, and not a single Fermi surface.
The zero-entropy degenerations that we discuss here remedy this problem.
\end{enumerate}

Using these insights, we make use of recently found black hole solutions,
which we suggest correspond to specific operators from a restricted subset of the CFT fields.
In the spirit of the operators discussed in ref. \cite{Berkooz2008},
we attempt to construct the Fermi surface operators while limiting ourselves
to a specific sector of $\mathcal{N}=4\ SU(N)$ SYM.
It has been shown (cf. (\cite{Beisert2004})) that this CFT has several sectors,
which contain operators mixing only with operators inside the sector.
The operators in these sectors, which are built from a specific subset of the theory's partons, are characterized by specific relations between their classical scaling dimension and charges.
Specifically, one of these sectors, the $PSU(1,1|2)$ sector, which we shall describe later in more detail, has interesting properties.
First of all, it contains a small number of partons, making it easier to consider all the relevant operators. Second, it has an outer $SU(2)$ symmetry rotating the two fermionic partons in the sector. Using this symmetry, it is possible to build operators which do not mix with any other operators, even within the sector.
Such operators have specific scaling relations between the charges, which we attempt to reproduce in the gravity theory. We have identified black holes in the gravity theory carrying the same charges as the $PSU(1,1|2)$ Fermi surface operators. These black holes can be degenerated to have zero entropy at extremality.
Using the conjectured duality, we can predict some of their properties from the CFT side
by using weak-coupling computations, which we can then try and verify in strong coupling in gravity.

The outline of the paper is as follows.
In Section \ref{res_summary} we briefly summarize our results concerning the quasinormal modes for an uncharged
massless scalar on the black hole background. Section \ref{black_hole_sec} describes in detail field theory construction of Fermi surface operators, the supergravity theory and the specific black hole background examined in this paper. In Section \ref{QNM_sec} we review the linear perturbation theory for a massless scalar, and describe
such a scalar in the supergravity theory. In Section \ref{radial_eq_sec} we solve the radial equation of motion at the near-extremal limit using a matched asymptotic expansion method, and Section \ref{angular_eq_sec} completes the analysis by solving the angular equation using an algebraic method which provides numerical solutions as well as some analytic results.
Finally, we show some explicit analytical and numerical results in Section \ref{results_sec}, and discuss them in Section \ref{discussion_sec}.

\section{Summary of Results}
\label{res_summary}
The black holes we are interested in are found within a consistent truncation of type $IIB$ Supergravity on $AdS_5 \times S^5$ which contains the graviton, three $U(1)$ vector fields and two uncharged massive scalars \cite{pope1,Cvetic:2005zi,Cvetic:2004ny,Mei2007, Chong2005,Chong2005b,Chong2007,Kunduri2006}.
We examine them in the limit where the black holes degenerate to zero-entropy configurations,
in which a ring of singularities is no longer cloaked by the horizon.
We review the conjecture that for a specific choice of charges these configurations are dual to Fermi Surface operators in the $PSU(1,1|2)$ sector of $\mathcal{N}=4$ $SU(N)$ $SYM$. This conjecture leads to predictions on the spectrum of quasinormal modes associated to these singularities which we attempt to verify.

For simplicity, we shall focus here on the spectrum of scalar quasinormal modes.
Using the symmetries of the metric, the problem of finding scalar QNMs reduces to a two dimensional PDE eigenvalue problem.
A further simplification occurs if the scalar field is massless, in which case
the linearized equations of motion can be reduced to a pair of coupled ODEs, an angular and a radial one,
which we shall attempt to solve.

It should be noted at this point that in order to compare the resulting spectrum to the spectrum
of the dual CFT, the black hole background together with the added scalar field should form
a consistent truncation of the complete gravitational theory, i.e. type $IIB$ theory.
In this work we highlight the only massless scalars which
can be added to the system such that together they form a consistent truncation.
Using the results in ref. \cite{Cvetic2000} we see that coupling the black hole to a dilaton and an axion yields a consistent truncation. Both of these are massless scalar fields. At the linearized level, they are both minimally coupled to the black hole metric.

For a generic choice of the black hole parameters, the separated equations of motion
are both Heun equations, second order regular differential equations with four regular singular points.
In the near-extremal limit, the distance between two of
the regular singular points in the radial equation goes to zero.
Using a matched asymptotic analysis described below, we are able to solve the
Heun's equation with two nearby singularities. The procedure is then applied to the radial equation.
The solution yields the separation constant $\lambda$ in terms of the complex frequency of the scalar, and we obtain
\begin{equation}
\omega= m_{\phi} \Omega_{\phi}-i 4\pi T \left(\frac{1}{2}\sqrt{1-\frac{\alpha \lambda(a,m_{\phi},\alpha)}{\alpha+4a+a^2 \alpha}} +\frac{1}{2}+n\right)
\end{equation}
expressed in terms of the temperature $T$ of the black hole, which approaches $0$ in the near-extremal limit.
Here, $a$ is the rotation parameter around one of the independent rotation planes, $m_{\phi}$ is the azimuthal quantum number associated with the scalar field mode, and $\Omega_{\phi}$ is the horizon's angular velocity. We have defined the parameter $\alpha= \frac{J_{\phi}/N^2}{(Q/N^2)^2}$, with $J_{\phi}$ the angular momentum and $Q$ the $U(1)$ charge. The AdS radius has been set to $1$ for convenience.  The positive integer $n$ is the integer number obtained from the radial equation.

Next, we use the Continued Fraction Method (CFM) to solve the
angular equation numerically for the separation constant,
which together with the above expression yields the quasinormal modes.
In addition, there are specific values of the parameters
where one is able to extract analytical solutions for the lowest eigenvalue.
The lowest eigenvalue for $\alpha = 2$, which is close to the $1/16 BPS$ bound,
and small enough azimuthal number $m_{\phi}$ is given by
\begin{equation}
\label{expression_omega_alpha_2}
\omega = m_{\phi}-i 4 \pi T \left(1+n+\left(\frac{m_{\phi }(1-a)}{2(1+a)}\right)\right).
\end{equation}
In addition, it is possible to extract analytic results for all values of $\alpha$ in the large-charge limit, $a \rightarrow 1$ limit,
when the black hole is rotating very rapidly.
We also discuss some of the properties of the eigenfunctions associated with these modes,
and compare them to known results obtained using the near-horizon techniques of Kerr/CFT.

In order to check if these results agree with our conjecture
for the dual of the black holes we recall that the construction
of the dual Fermi surface was made in the large charge regime, i.e $a \rightarrow 1$.
Taking this limit in equation
(\ref{expression_omega_alpha_2})
we find that these quasinormal modes look like those of the BTZ black hole
    \begin{align}
    \omega_n^{(L)}= & k-i4\pi T_L \left(h_L+n\right) \\
    \omega_n^{(R)}= & -k-i4\pi T_R \left(h_R+n\right).
    \end{align}
It is well known that the BTZ black hole is dual to a $1+1$ CFT and that the quasinormal modes above are the poles of the retarded Greens function of an operator with dimension $(h_L,h_R)$ in this CFT. Therefore, our result (\ref{expression_omega_alpha_2}) implies that the singular degenerate black hole, together with the perturbation we have considered, are dual to the left sector of a $1+1$ CFT.
Furthermore, the dimension of the
operator corresponding to the perturbation is equal to $1$,
which is compatible with a free fermion bilinear.
Actually, the $PSU(1,1|2)$ sector can be thought of as a $1+1$ CFT with the coordinate $\phi$ playing the role of $x$ and thus $m_{\phi}$ the role of $k$.
Thus, these findings lend support to the conjecture that the degeneration of a charged rotating black hole considered in this paper is dual to a Fermi surface construction in the $PSU(1,1|2)$ sector.

\section{Extremal Black Holes and Fermi Surfaces}
\label{black_hole_sec}

Several works \cite{Lee:2008xf,Rey:2008zz,cubrovic:2009ye,Liu:2009dm,Faulkner:2009am,Faulkner:2010zz,Faulkner2011b,Sachdev2010,Iizuka2011,DeWolfe2012} have suggested that extremal charged black holes are dual to Fermi surfaces on the CFT side. In these studies the authors compute the Green's function of Fermionic composite operators via a fermionic propagator in the presence of the black hole. They find sharp features in these functions at finite momentum and small frequency, which is an indication for the presence of a Fermi surface. At the same time, some of these features also suggest these are exotic Fermi surfaces and in particular non-Fermi liquids.
In addition, it has been problematic to establish that these
are indeed Fermi surfaces on the field theory side since in many cases the field theory dual is not known.
For a detailed discussion of the subject see \cite{Faulkner2011} and references therein.

One of the subtle issues associated with this construction has to do with the entropy.
The black holes used have finite entropy at zero temperature, whereas a conventional CM Fermi surface ground state would have zero entropy at zero temperature. This might be an effect of large-N versus finite-N, since CM systems are finite-N systems, but in any case it implies that the phenomenology of these black holes as Fermi surfaces would be different than that of known Fermi surfaces in CM systems.

We are thus led to consider, on the gravity side, zero-entropy extremal black holes.
One of the ways to achieve such a zero-entropy construction is to
degenerate an extremal black hole into a zero-entropy singularity.
Such an object would correspond to a system with zero entropy at zero temperature, at least in the large-N limit.
Extremal black holes and their degenerations have been studied in various supergravity theories.
The most studied solutions are in $AdS_5\times S^5$ and the dual $\mathcal{N}=4$ SYM
\cite{Chow:2008dp,Cvetic:2005zi,Cvetic:2004ny,Mei2007}.
Similar constructions were examined in \cite{Dias2007}
for a different type of compactification, $T^4 \times S^1$.
There, the relation between zero-entropy gravity
configurations and Fermi surfaces in the low-energy
dual CFT was employed in order to study the phenomenon of superradiance.

Recently, Extremal black holes with Vanishing Horizon (EVH) have been studied in detail in several works \cite{Fareghbal2008, Sheikh-Jabbari2011, Boer2011}\footnote{We thank M.M. Sheikh Jabbari for bringing these works to our attention.}, both from the gravity and field theory point of view.
It has been shown that their near-horizon is typically a pinched $AdS_3$, which becomes a pinched BTZ for near-extremal black holes.
It will be interesting to examine our results in light of these works.

Our goal is therefore twofold. First, we wish to identify a Fermi surface-like state in the CFT.
Next, we would like to study its gravitational dual and verify that its extremal limit indeed
approaches a zero-entropy configuration.
We do the former in \ref{fermi_surface_subsec}, and introduce our candidate black hole solution in \ref{black_hole_subsec}.
The rest of the paper is devoted to studying the quasinormal modes of a scalar in the background of the black hole and to the verification that they exhibit the expected behavior.

\subsection{Fermi Surface Operators and the \it{PSU(1,1$|$2)} Sector}
\label{fermi_surface_subsec}
We would like to construct the Fermi surface within the context of $\mathcal{N}=4$ $SU(N)$ SYM.
There have been some attempts at constructing such Fermi surface operators from the partons of $\mathcal{N}=4$ SYM and finding black hole solutions which reproduce some of their properties \cite{Berkooz2007, Berkooz2008}.
However, constructing a state using the full field content of $\mathcal{N}=4$ is difficult,
since the theory contains several scalar fields, which usually trigger instabilities.
Even if no instabilities exist, the scalar fields tend to mix with the fermions, potentially increasing the entropy.

Fortunately, there exist several subsets, or sectors, of the theory's partons mixing only among themselves under the dynamics of the theory.
The subsectors of $\mathcal{N}=4$ have been studied extensively, and have been classified in \cite{Beisert2004}.
A sector which seems especially suited to our purposes,
is the $PSU(1,1|2)$ sector, which contains only two fermions and two scalars.
This sector has been analyzed in great detail in ref. \cite{Zweibel2006, beisert2007,Zwiebel2008}.
Within the $PSU(1,1|2)$ sector, one can build Fermi surface like
operators which are eigenstates of the dilatation operator, i.e. they do not mix with any other operators.
We will consider an $SU(4)$ rotated version of this sector as it appeared in ref. \cite{Beisert2004}.

The $PSU(1,1|2)$ sector contains only partons with the following relations between the charges
\begin{equation}
\label{psu_constraint}
\Delta_0 = 2 J_L + \hat{Q}_1 + \hat{Q}_2 + \hat{Q}_3 = 2J_R + \hat{Q}_1 + \hat{Q}_2 - \hat{Q}_3
\end{equation}
where $\Delta_0$ is the classical scaling dimension, $J_L$ and $J_R$ are the $SU(2) \times SU(2)$ quantum numbers, and
$\hat{Q}_1$, $\hat{Q}_2$ and $\hat{Q}_3$ are the $SU(4)$ $R$ charges spanning the $SO(2)\times SO(2) \times SO(2)$ Cartan subalgebra of $SO(6)\cong SU(4)$.

The constraints (\ref{psu_constraint}) retain only four types of partons from the full set
\begin{align}
\phi_1^{(k)} &\equiv \mathcal{D}^k_{1\dot{1}} \phi_{24} & \phi_2^{(k)} \equiv \mathcal{D}^k_{1\dot{1}} \phi_{34} \nonumber \\
\psi^{(k)} &\equiv \mathcal{D}^k_{1\dot{1}} \psi_{14} & \bar{\psi}^{(k)} \equiv \mathcal{D}^k_{1\dot{1}} \bar{\psi}^1_1.
\end{align}
Here, $D_{1\dot{1}}$ is the covariant derivative $D_{\alpha \dot{\alpha}}$ with $\alpha = \dot{\alpha} = 1$.
The symmetry inherited from the full theory is $PSU(1,1|2)$.
In addition there is an $SU(2)$ automorphism under which both $\phi_1^{(k)}$ and
$\phi_2^{(k)}$ are singlets for all $k$, and for each $k$, $\psi^{(k)}$ and $\bar{\psi}^{(k)}$ are a doublet.
This symmetry exists only within the sector, meaning it is not inherited from the full theory.
Following the notations of ref. \cite{beisert2007}, we
use indices in order to denote the accidental $SU(2)$ symmetry, defining
\begin{align}
\psi^{(k)} \equiv \psi_>^{k} && \bar{\psi}^{(k)} \equiv \psi_<^{k}.
\end{align}

While highly restricted, working with the full $PSU(1,1|2)$ sector can still be problematic.
There are two scalars in the sector, which may cause instabilities and produce large mixing effects.
In order to restrict their mixing, we use the outer $SU(2)$ automorphism, which we will refer to as
pseudospin, denoted $SU(2)_p$.
From the charges listed above, one can easily see that operators which contain $\psi_<^{k} \psi_>^{j}$
will generally mix with operators such as $\phi_1 \phi_2^{(j+k+1)}$.
However, mixing occurs only through the singlet channel of the pseudospin,
i.e. if the fermion bilinear is taken to be in the triplet of the pseudospin, it will not mix with
the scalar bilinear.
Thus, by using fully symmetric combinations in the indices $\{<,>\}$,
we can build Fermi surface operators out of the $\psi$'s which do not mix with operators built from scalar partons.

The simplest such construction is obtained if we build the operators from a single fermion $\psi_>^{k}$. In fact the collection of $D_{1\dot{1}}^m\psi_>^{k}$ with any number of derivatives $m$, comprises a closed subsector as well, namely the fermionic $SU(1,1)$ sector discussed in \cite{beisert2007} (there it is called the fermionic $sl(2)$ sector). This can be shown either using the oscillator formalism or, in the notation we have set above, by considering the pseudospin symmetry.
The single type of parton within this fermionic sector allows one to easily construct a Fermi surface operator. We review this construction explicitly below.

This operator does not mix with any other operators in the theory, and is thus an eigenstate of the dilatation operator. Such an operator, containing derivatives of the fermion ranging from $0$ to some large $K$, was studied in ref. \cite{Berkooz2007,Berkooz2008}. Interestingly, it was found in ref. \cite{Berkooz2008} that the dimension of such operators, both in weak and strong coupling, is the classical dimension with corrections of order of the inverse of the (large) charge.

The next step is to compute the charges of these states (see eq. (\ref{charges_eq}) for a similar computation),
and look for black-hole like states with the same charges.
It turns out, however, that they do not have finite entropy,
as noted in ref. \cite{Berkooz2008}. Therefore, as mentioned earlier, they do not make ideal candidates for studying CM Fermi surfaces.
Nevertheless, both in weak and strong coupling the dimensions are the classical dimensions with corrections of order of the inverse of the (large) charge.

One can instead use the $SU(2)_p$ to rotate the above operator to a different state in the same total  $SU(2)_p$ multiplet.  Such an operator will have both $\psi_{<}$ and $\psi_{>}$ and thus also different charges. One can then hope to find a black hole with the appropriate charges which has zero entropy in the zero temperature limit. We describe below the construction of such an operator and the corresponding black hole,
which can be degenerated to zero entropy in the extremal limit.

\subsubsection{Explicit Construction}
\label{Construction of Fermi surface subsubsection}
The quantum numbers of $\psi_>^{n}$ and $\psi_<^{n}$ play a central role in our construction. Thus, to preserve the clarity of the discussion, we add a tabulation of them.
\begin{table}[H]
\centering
\begin{tabular}{||c||c|c|c||c|c||c|c|c||}
  \hline
  \hline
   & $j_L$ & $j_R$ & $Q_1$ & $Q_2$ & $Q_3$ & $j_{p}$ \\
   \hline
  $\psi_>^{n}$ & $\left(n+1\right)/2$ & n/2 & 1/2 & 1/2 & 1/2 & 1/2  \\
  $\psi_<^{n}$ & n/2 & $\left(n+1\right)/2$ & 1/2 & 1/2 & -1/2 & -1/2  \\
 \hline
  \hline
\end{tabular}
\caption{Quantum numbers for the fermionic content of the \textit{PSU(1,1$|$2)} subsector, $j_{p}$ refers to the angular momentum of the $SU(2)_p$ automorphism.}
\end{table}

Note that each of these fields also sits in the adjoint of the gauged $SU(N)$ so it has an additional adjoint index which we have so far suppressed. We will sometimes wish to write this index explicitly denoting the field in the following way $\psi_{>(j)}^{n}$ where $j=1...g$ and $g=N^2-1$.

We will begin by constructing the operator discussed in \cite{Berkooz2008} and then present its rotated version, which is the one relevant to the black hole we consider below. The operator is given by
\begin{equation}
\label{first_operator}
\mathcal{O}_1^{(K)} = \prod_{j=0}^{2K} \psi_{>(1)}^{j}\psi_{>(2)}^{j}\cdots\psi_{>(g)}^{j}\equiv \prod_{j=0}^{2K} \epsilon^{a_1 a_2 ... a_g}\psi_{>(a_1)}^{j}\psi_{>(a_2)}^{j}\cdots\psi_{>(a_g)}^{j}.
\end{equation}
Each term in the product is gauge invariant by itself as can be seen from the right most equality.
To avoid cumbersome notations we will denote
\begin{align}
Jdet[\Psi] = \epsilon^{a_1 a_2 ... a_g} \Psi_{(a_1)} \Psi_{(a_2)} ... \Psi_{(a_g)} && \Psi = \sum_{a=1}^{g} \Psi_{(j)} T^a\ (g=dimG).
\end{align}
The operator described in eq. \ref{first_operator} then reads
\begin{equation}
\mathcal{O}_1^{(K)}=\prod_{j=0}^{2K}Jdet[\psi_{>}^{j}].
\end{equation}

Proceeding, we construct the rotated Fermi surface operator with an equal number of $\psi_>$ and $\psi_<$,
\begin{equation}
\label{operator}
\mathcal{O}^{(K)} = Sym\left[\prod_{j=0}^{K-1} Jdet\left[\psi_<^{j}\right] \prod_{m=K}^{2K-1} Jdet\left[\psi_>^{m}\right]\right]
\end{equation}
where $Sym[\text{  }]$ stands for a symmetrization of the operator with respect to the $\{<,>\}$ indices,
placing the operator in the highest $SU(2)_p$ state, with $J^2 = 2K N^2(2K N^2+1)$ and $J_z=0$.

The operator we are considering is not a BPS operator so there is no a priori reason for its dimension in the strongly coupled theory to be close to the classical one. However, motivated by \cite{Berkooz2008}, we might hope that the corrections to the dimension scale as $O(\frac{1}{K})$. Therefore we require a large charge limit, $K >> 1$ where we may have some control over the corrections.
One can easily compute the charges for such an operator
\begin{align}
\label{charges_eq}
\hat{Q}_1 &= \hat{Q}_2 \equiv \hat{Q} = N^2 K \nonumber \\
\hat{Q}_3 &= 0 \nonumber \\
J_L &= N^2 \left(\sum_{j=0}^{K-1}\frac{j+1}{2} + \sum_{m=K}^{2K-1}\frac{m}{2}\right) = N^2 K^2 \nonumber \\
J_R &= N^2 \left(\sum_{j=0}^{K-1}\frac{j}{2} + \sum_{m=K}^{2K-1}\frac{m+1}{2}\right) = N^2 K^2 = J_L \nonumber \\
\Delta_0 &= 2J_R + \hat{Q}_1 + \hat{Q}_2 = N^2(2K^2 + 2K)
\end{align}
with $\Delta_0$ the classical scaling dimension.

One could also construct similar states with higher angular momentum by leaving unfilled 'holes' in the  operator (\ref{operator}). We are still interested in the highest outer $SU(2)$ state so it should be symmetrized with respect to $\{<,>\}$.
In order for the symmetrized state not to vanish, any $\psi_>$ must appear with a different number of derivatives compared to any $\psi<$ before symmetrization.
For example, adding one derivative to the last operator in the sum, and creating the operator
\begin{equation}
\mathcal{\tilde{O}} =
Sym\left[\prod_{j=0}^{K-1} Jdet\left[\psi^{<j}\right] \prod_{m=K}^{2K-2} Jdet\left[\psi^{>m}\right] Jdet\left[\psi^{>2K}\right] \right].
\end{equation}
Such 'sparse' operators can only mix with other sparse states with the same
number of partons, provided they have the same angular momentum, i.e.
that the total number of derivatives, which are distributed among the fermionic operators, sum up to the same number.

These operators satisfy the scaling relation
\begin{equation}
\alpha \equiv \frac{\frac{(J_L + J_R)}{N^2}}{\left(\frac{\hat{Q}}{N^2}\right)^2}
\end{equation}
where for the operator (\ref{operator}) one gets $\alpha=2$ while the sparse operators have $\alpha = 2 + \frac{1}{2K^2}$.

Motivated by these results, in Section \ref{black_hole_subsec} and onward, we will focus on extremal black holes with $\hat{Q}_1=\hat{Q}_2$, $\hat{Q}_3=0$ , $J_L=J_R$ which also satisfy the scaling relations \ref{charges_eq}. As a consistency check we will verify that they indeed have $E=\Delta_0$ up to $O\left(\frac{1}{K}\right)$.

\subsection{Description of the Black Hole Solution}
\label{black_hole_subsec}
We work with a consistent truncation of type $IIB$ Supergravity on $AdS_5 \times S^5$ described in ref. \cite{cvetic}.
The field content comprises of the metric, two neutral scalars and three abelian $U(1)$ fields.
It can also be consistently coupled to a dilaton and an axion, as we shall discuss later.
The action of the bosonic part of the supergravity is given by
\begin{align}
\label{lagrangian}
S = &\int d^5 x \sqrt{-g} \left[ R - \frac{1}{2} \sum_{\alpha=1}^{2}{(\partial \varphi_{\alpha})^2} + \sum_{i=1}^{3}{\left(4 l^{-2} X_i^{-1} - \frac{1}{4} X_i^{-2} \mathcal{F}^i_{\mu \nu} \mathcal{F}^{i \mu \nu}\right)}\right] + \nonumber \\
&\int d^5 x \frac{1}{24} |\epsilon_{ijk}| \epsilon^{uv \rho \sigma \lambda} \mathcal{F}^i_{uv} \mathcal{F}^j_{\rho \sigma} A_{\lambda}^k.
\end{align}
Here,  $l_{ads} \equiv l$ is the AdS radius, $A^i$ are the three $U(1)$ gauge fields, and
$X_i$ are the three uncharged scalars, constrained by $X_1 X_2 X_3 = 1$. These scalars may
be parameterized by
\begin{align}
X_1 = e^{-\frac{1}{\sqrt{6}}\phi_1 -\frac{1}{\sqrt{2}} \phi_2},\ \ \ \  & X_2 = e^{-\frac{1}{\sqrt{6}}\phi_1 +\frac{1}{\sqrt{2}} \phi_2},
& X_3 = e^{\frac{2}{\sqrt{6}}\phi_1}.
\end{align}

We will use black hole solutions in five dimensions that first appeared in \cite{pope1},
while following the notations of \cite{Mei2007} which contains a generalization of these solutions.
The black hole presented here is rotating around the two independent directions,
$\psi$ and $\phi$, with angular momenta $J_{\psi}$ and $J_{\phi}$.
In terms of the field-theory angular momenta $J_L$ and $J_R$ these are
\begin{align}
\label{angular momentum relation}
J_{\phi}=J_L + J_R && J_{\psi}=J_L - J_R.
\end{align}
This solution places no restrictions on the values of $J_{\psi}$ and $J_{\phi}$.
The black hole is also charged under the three $U(1)$'s. The relation between these charges and the field-theory charges is
\begin{equation}
Q_i=\frac{\hat{Q}_i}{l}.
\end{equation}
The charges $J_{\phi}$, $J_{\psi}$, $Q_1(=Q_2)$, $Q_3$ and $E$ are parameterized by $\delta_1$, $\delta_3$, $m$, $a$, $b$, in the conventions of \cite{Mei2007}. Our $l$ is denoted there by $1/g$.

The near-horizon limit of these black holes at extremality was analyzed in \cite{Chow:2008dp} in the context of Kerr/CFT.
The case of three equal charges was studied in greater detail \cite{Davis2005, Wu2009,Birkandan2011}.
Quasinormal modes when all three charges are equal were modes explicitly in \cite{Aliev2008}.

Restricting ourselves to
$Q_3 = 0$ and $J_L-J_R = J_\psi = 0$ amounts to setting $b = 0$ and $\delta_3 = 0$.
The solution then reduces to the following metric
\begin{equation}
\begin{split}	
\label{metric2}
ds^2 = H_1^{2/3} \Biggl\{ &  \left(x^2 + y^2\right) \left(\frac{dx^2}{X} + \frac{dy^2}{Y} \right) -
			\frac{ X \left( dt - y^2 d \sigma \right)^2}{\left(x^2 + y^2\right) H_1^2} + \Biggr. \\
			\Biggl.& \frac{Y \left[dt + \left(x^2 + 2 m s_1^2\right) d\sigma \right]^2}{\left(x^2 + y^2\right) H_1^2}  + y^2 x^2 d\chi^2 \Biggr\}
\end{split}
\end{equation}
where we define $c_1 = cosh\delta_1$,$s_1 = sinh\delta_1$ and $\Sigma_a = 1-a^2/l^2$.
The functions used in the metric are
\begin{equation}
\begin{split}
X &= -2m + \left(a^2 + x^2\right) + l^{-2}\left(a^2 + 2m s_1^2 +x^2\right) \left( 2m s_1^2 + x^2 \right) \\
Y &= \left(a^2 - y^2\right) \left( 1 - l^{-2} y^2\right)  \\
H_1 &= 1 + \frac{2 m s_1^2}{x^2+y^2}.
\end{split}
\end{equation}
In addition, the gauge field and scalar backgrounds are
\begin{equation}
\begin{split}
A^1 &= A^2 = \frac{2 m s_1 c_1 (dt - y^2 d\sigma)}{(x^2 + y^2)H_1} \\
A^3 &= \frac{2 m s_1^2 y^2 d\chi}{(x^2+y^2)}  \\
X_1 &= X_2 = H_1^{-1/3} \\
X_3 &= H_1^{2/3}.
\end{split}
\end{equation}
One is often interested in the coordinate frame which is non-rotating at infinity
(denoted by $\tau$, $\phi$, $\psi$ for the time and the two azimuthal coordinates, respectively),
since the CFT quantities are defined in this frame.
In the specific case we are examining,
the linear transformation between the different coordinate sets reduces to
\begin{equation}
\begin{split}
t &= \frac{1}{\Sigma_a} \tau - \frac{a}{\Sigma_a} \phi,  \\
\sigma &= \frac{1}{l^2 \Sigma_a} \tau - \frac{1}{a \Sigma_a} \phi, \\
\chi &= \frac{1}{a} \psi.
\label{angular_trans}
\end{split}
\end{equation}
Finally, using the transformation
\begin{equation}
\begin{split}
\label{x_to_rho}
& x^2 =r^2-\frac{4}{3} m s_1^2 \\
& y^2=a^2 cos^2\theta
\end{split}
\end{equation}
in the non rotating frame, the metric explicitly takes the form of a black hole with an horizon of $S^3$ topology located at $r=r_0$,
which is the largest root of $\Delta(r)=\frac{x^2 X}{r^2}$.\\
The Ads/CFT correspondence suggests that the following relation holds \cite{Aharony2000}:
\begin{equation}
\frac{\pi l_{ads}^3}{4G_5} = \frac{N^2}{2}.
\end{equation}
Using this relation, the black hole's global charges and thermodynamic quantities are expressed by
\begin{equation}
\begin{split}
J_{\phi} &= \frac{\pi}{4 G_5} \frac{2 m a \left(1+ s_1^2\right)} {{E_{a}}^2} =N^2 l^{-3} \frac{ m a \left(1+ s_1^2\right)} {{E_{a}}^2} \\
Q_{1} &= Q_{2} \equiv Q = \frac{\pi}{4 G_5} \frac{2 m s_{1} c_{1}}{E_{a}}=N^2 l^{-3} \frac{ m s_{1} c_{1}}{ E_{a}} \\
E &= \frac{\pi}{4 G_5} \frac{m [ (2(l^{-4} a^4 + \Sigma_a + 1) + l^{-2} a^2 (\Sigma_a -2))s_1^2 + \Sigma_a  + 2] }{\Sigma_a^2} \\
S &= \frac{\pi^2}{4 G_5} \frac{2 \sqrt{r_0^2 - \frac{4}{3} m s_1^2} (a^2 + \frac{2}{3} m s_1^2 + r_0^2)}{\Sigma_a}=N^2 l^{-3} \pi \frac{ \sqrt{r_0^2 - \frac{4}{3} m s_1^2} (a^2 + \frac{2}{3} m s_1^2 + r_0^2)}{\Sigma_a} \\
T &= \frac{r_0\left(\partial _r\Delta \right)\sqrt{r_0^2-\frac{4}{3}m s_1{}^2}}
     {4\pi \left(a^2+r_0^2+\frac{2}{3}m s_1{}^2\right)\left(r_0^2-\frac{4}{3}m s_1{}^2\right)}.
\end{split}
\end{equation}
Here, $r_0$ is again the radius at which the event horizon is located.
Finally, the angular velocity around the $\phi$ direction at the event horizon,
with respect to an asymptotically non-rotating AdS space, is given by
\begin{equation}
\Omega_{\phi} = \frac{a\left(1+l^{-2}\left(r_0 + \frac{2 m s_1^2}{3}\right)\right)}{a^2+r_0^2 + \frac{2 m s_1^2}{3}}.
\end{equation}

\subsubsection{The extremal limit of the black hole}
In this paper we focus on the black hole very close to the extremal limit.
Its temperature is thus very close to zero.
To examine this case we need to find the location of the horizon which is given by $\Delta = 0$.
There are two solutions for $\Delta = 0$, given by $X = 0$ or $x^2 = 0$.
The solutions to these equations are:
\begin{subequations}
\begin{align}
x^2=0:\  & r_1^2 = \frac{4m s_1^2}{3} \\
X =0:\  &r_2^2 = \frac{-1 - a^2 - \frac{4 m s_1^2}{3} + \sqrt{(a^2-1)^2 + 8m (1+s_1^2)}}{2} \\
       &r_3^2 = \frac{-1 - a^2 - \frac{4 m s_1^2}{3} - \sqrt{(a^2-1)^2 + 8m (1+s_1^2)}}{2} < 0.
\end{align}
\end{subequations}
Here, and for the rest of this paper, we set $l=1$.
Factors of $l$ can be restored using dimensional analysis.
In this notation we have:
\begin{equation}
\label{large_X}
X=(r^2-r_2^2)(r^2-r_3^2)=(x^2 + (r_1^2-r_2^2))(x^2 + (r_1^2-r_3^2)) \equiv (z-z_2)(z-z_1).
\end{equation}
There are two horizons $r_1$ and $r_2$ and the extremal limit is obtained when $r_1=r_2$ in which case we find that the entropy is zero and the solution is not a real black hole. This is also the case for $r_1>r_2$ since $\chi$ circles contract at the radius $r_1$. Only for $z_2\equiv r_2-r_1>0$ we get a real black hole and we can use $z_2$ as a measure of the distance of the black hole from an extremal one.

Note, that the solution has an ring of singularities at $(t,x,y,\phi,\psi)=(t,0,0,\phi,\psi)$ which is hidden behind the horizon above extremality, for $z_2>0$, and becomes naked when $z_2 = 0$. Moreover, it turns out that the distance to this singularity, i.e the distance from a finite radial position to the horizon in the submanifold $y=0$, is finite, even as the limit of $z_2\rightarrow0$ is taken.
Hence, there is no standard near-horizon extremal limit.

It will be convenient to use a parametrization in terms of $\alpha= \frac{J_{\phi}/N^2}{(Q/N^2)^2}$ and $z_2$.
The old parameters in terms of $\alpha$ and $z_2$ are
\begin{equation}
\begin{split}
m &= \frac{\left(a^2+z_2\right) \left(1+z_2\right)}{2}+\frac{2 a^2}{\alpha^2}+\frac{a^3+2 a z_2}{\alpha} \\
m s_1^2 &= \frac{a}{\alpha} \\
z_1 &=  -\frac{4 a+\left(a^2+1+z_2\right) \alpha }{\alpha}.
\end{split}
\end{equation}
The BPS limit of the black hole is
\begin{equation}
E=J_{\phi}+J_{\psi}+Q_1+Q_2+Q_3=J_{\phi}+2Q_1
\end{equation}
and it is satisfied for $z_2=0$ (since the black hole must also be extremal) and $\alpha=2$.
Coming back to the issue of superradiance, we would expect qualitatively to have superradiance and therefore also an instability if $\Omega_{\phi}>1$. Rewriting $\Omega_{\phi}$ in terms of $\alpha$ and $z_2$ we get
\begin{equation}
\Omega_{\phi}=\frac{\alpha+2a+z_2}{a \alpha+2+z_2}.
\end{equation}
Thus, in the extremal limit, $\Omega_{\phi}>1$ is equivalent to $\alpha > 2$ and for the BPS black hole $\Omega_{\phi}=1$. Therefore, for a black hole very close to being extremal, superradiant instability should appear a little above $\alpha=2$.
One should note that for any $z_2$ and $\alpha$, if $a=1$ then $\omega_{\phi}=1$.
It is illuminating to  rewrite the temperature and entropy of the black hole in terms of $z_2$ and $\alpha$:
\begin{equation}
\begin{split}
S &= N^2 \frac{2\pi \sqrt{z_2} \left(a^2+\frac{2 a}{\alpha }+z_2\right)}{\Sigma_a} \\
T &= \frac{\sqrt{z_2} \left(4 a +\alpha + a^2 \alpha +2 \alpha z_2\right)}{4 a \pi +2 \pi \alpha   \left(a^2+z_2\right)}.
\end{split}
\end{equation}
Our main interest are rapidly rotating near extremal black holes which means we are considering $\Sigma_a<<1$ and $z_2<<1$. From the expression for the entropy it is clear that the relative size of these quantities is quite important.
In particular, in this paper we work in the limit $\sqrt{z_2}<< \Sigma_a$ which implies vanishing entropy.
This case seems more relevant to a Fermi surface description.
We will analyze the opposite case, for which the entropy is finite and large in an upcoming paper \cite{work_in_progress}.

\section{Scalar Perturbations and Quasinormal Modes}
\label{QNM_sec}
\subsection{Quasinormal Modes of Linear Perturbations}
There are several types of instabilities of black holes.
The simplest one has to do with linearized stability about the black hole background.
This will be the focus of our paper.
The issue of thermodynamic instability and global instability to
decay into other phases will be discussed in a separate paper \cite{work_in_progress}.

In principle the linear analysis is straightforward, although sometimes involved -
one writes down, and solves, the
linearized equations of motion around the black hole background for the different fields present in the (super)gravity theory.
To find the energy spectrum of the field on this background, one uses the ansatz $\Psi(x) = e^{-i \omega \tau} \tilde{\Psi}(\vec{x})$.
In this convention,
placing suitable boundary conditions at the horizon and at infinity implies a quantization of $\omega$'s. The spectrum of $\omega$'s and the associated
modes are called QuasiNormal Modes (QNMs for short).
For a review of QNMs in black hole physics see ref. \cite{Berti2009}.
Since the black hole is dissipative, most $\omega$-s would have an imaginary part
and instability is associated with a positive imaginary part of $\omega$.

Although one can write the equations and attempt to solve them numerically,
this is usually a complicated task. It is also difficult to obtain analytical results.
This is particularly true in AdS background where the change in asymptotic
behavior introduces more singular points (in the technical sense of differential equations) into the equations of motion.
Analytical results can be obtained only if the equations are separable, which happens only in special cases.
This is also the case in which the numerical approach, perhaps combined with partial analytic results, is practical.

In addition to being important quantities, characterizing different gravitational backgrounds, and specifically their stability,
one can also hope to use QNMs in the context of the AdS/CFT correspondence.
Roughly speaking, black holes are dual to thermal states in the gauge theory, and quasinormal modes would then correspond to decay times of different types of excitations.
However, in order to make use of the full power of the duality, the perturbations studied on the gravity side
must be part of a consistent reduction of type $IIB$ supergravity on $AdS_5 \times S^5$.
This compounds the problem since the equations of motion for a given set of fields in type IIB may or may not factorize,
and we do not have the freedom of changing their couplings.

\subsection{A Consistent Truncation of Type IIB Including a Massless Scalar}
The full spectrum of the Kaluza-Klein compactification of type IIB Supergravity on $S^5$ was found in refs. \cite{Kim:1985ez, Gunaydin:1984fk}.
Its truncation to the massless sector is a consistent truncation and results in five dimensional $\mathcal{N}=8$ gauged $SO(6)$ Supergravity \cite{Gunaydin:1984fk, Gunaydin:1985cu}.
The latter five dimensional theory consists of gravity, $15$ gauge fields in the adjoint of $SO(6)$, twelve 2-form gauge potentials in the $6$ and $\bar{6}$ of $SO(6)$, 42 scalars in the $1+1+20'+10+\bar{10}$ representations of $SO(6)$ (the  20' is spanned by symmetric unimodular matrices) and their fermionic superpartners. This theory has, in addition to the $SO(6)$ gauge symmetry, a global $SL(2,\mathds{R})$ symmetry inherited from the ten type IIB dimensional Supergravity.

A smaller consistent truncation can be constructed by restricting the above theory to the singlets of $U(1)^3 \times SL(2,\mathds{R})$ subgroup, where $U(1)^3$ is the Cartan subgroup of $SO(6)$. This amounts to keeping only 2 scalars out of the $20'$ and the three $U(1)$ gauge fields in the Cartan subgroup. The resulting theory is an $\mathcal{N}=2$ supergravity with the Lagrangian (\ref{lagrangian}). This theory was shown to be a consistent truncation of type IIB Supergravity in \cite{cvetic} by an explicit construction of the full nonlinear Kaluza-Klein ansatz. The black hole described above is a solution of this theory.

We are interested in finding a field which can be used to perturb the black hole background. In addition to keeping its perturbation equations simple, we require that, when added to the above theory, it forms a consistent truncation of type IIB Supergravity. Indeed, relaxing the condition for $SL(2,\mathds{R})$ invariance one gets the previous truncation with the addition of the axion $\chi$ and dilaton $\phi$ (the scalars in the $1+1$), and the following addition to the Lagrangian (\ref{lagrangian})
\begin{equation}
\label{lag_add}
\delta \mathfrak{L}=\sqrt{-g}\left(-\frac{1}{2}\partial_{\mu}\phi \partial^{\mu}\phi-\frac{1}{2} e^{2\phi}\partial_{\mu}\chi \partial^{\mu}\chi\right).
\end{equation}
As noted in ref. \cite{Cvetic2000} this is also a consistent truncation of the ten dimensional theory.
Indeed, this is the only single massless scalar enlargement which forms a consistent truncation.
We cannot change the lagrangian (\ref{lagrangian}) + (\ref{lag_add}),
but remarkably the equations of motion for the axion and
dilaton are separable at the linearized level for the black hole in question.

\subsection{Separating the Linearized Equations of Motion}
The equation of motion for a minimally coupled massless uncharged scalar in the black hole background
is given by
\begin{equation}
\label{EOMscalar}
\begin{split}
&\frac{1}{x} \partial_x \left(x X \partial_x \Phi \right) +
\frac{1}{y} \partial_y \left(y Y \partial_y \Phi \right) + \frac{x^2 + y^2}{x^2 y^2} \partial_{\chi}^2 \Phi +\\
&\frac{1}{Y} \left(y^2 \partial_t + \partial_{\sigma} \right)^2 \Phi -
        \frac{1}{X} \left(\left(x^2 + 2 m s_1^2\right) \partial_t - \partial_{\sigma} \right)^2 \Phi = 0.
\end{split}
\end{equation}
We carry out the separation of variables using the ansatz
\begin{equation}
\Phi = e^{-i w \tau + i m_{\phi} \phi +i m_{\psi} \psi} R(x) S(y)
\end{equation}
written in terms of the quantum numbers
of the asymptotically non-rotating coordinates, $\tau$, $\phi$, $\psi$.
Rewriting (\ref{EOMscalar}) in terms of the coordinates appearing in the equation of motion we get
$\Phi = e^{i a m_{\psi } \chi +i t \left(a m_{\phi}-\omega \right)+i \sigma  \left(-a m_{\phi}+a^2 \omega \right)} R(x) S(y)$.
Plugging into the equation of motion (\ref{EOMscalar}), and separating variables, we obtain two ordinary differential equations. The radial equation is given by
\begin{equation}
-\frac{\frac{1}{x }\partial _x\left(xX\partial _xR(x)\right)}{R(x)}+\frac{a^2 m_{\psi }{}^2}{x^2}-\frac{\left(\left(x^2+2m s_1^2\right)\left(-\omega + a m_{\phi }\right)+a m_{\phi } -a^2\omega \right){}^2}{X}=\lambda,
\label{radial_eq}
\end{equation}
while the angular equation assumes the form
\begin{equation}
\frac{\frac{1}{ y }\partial _y\left(y Y \partial _yS(y)\right)}{S(y)}-\frac{a^2 m_{\psi }{}^2}{y^2 }-\frac{\left(y^2 \left(a m_{\phi }- \omega \right)-am_{\phi }+a^2\omega \right){}^2}{Y} = \lambda.
\end{equation}

Motivated by the conjectured field-theory duals, we are interested in near extremal black holes which are rotating rapidly. This translates to the limits
\begin{align}
z_2 &\rightarrow 0 \nonumber \\
a &\rightarrow 1
\end{align}
where $z_2$ is defined in (\ref{large_X}) and $a$ parameterizes rotation around $\phi$, as seen in the definition of the metric. The limit $a \rightarrow 1$ implies fast rotation, and $z_2 \rightarrow$ implies extremality.

In the following sections, we solve this equation in the limit
$\sqrt{z_2} << 1-\frac{1}{a^2}$,
so the black holes being studied are arbitrarily close to extremality.
In Section \ref{radial_eq_sec} we solve the radial equation analytically with the additional limit $\omega = m_{\phi} \Omega_{\phi} + \sqrt{z_2} \Delta$,
using matched asymptotic expansion.
We obtain $\lambda$ as a function of $\omega$ and a new integer $n$.
In Section \ref{angular_eq_sec} we solve the angular equation in this limit and find the quantized values of $\lambda$. Comparing these with the
results from the radial equation yields the QNM spectrum in this limit.

\section{The Radial Equation}
\label{radial_eq_sec}
We begin the analysis with the radial equation (\ref{radial_eq}). For the solution to be normalizable, we impose a boundary condition at $\infty$, $R(x)\rightarrow 0$. At the horizon we demand that there would be no outgoing waves, since we are in the classical regime.
As we show in subsection \ref{radial_heun} below, the radial equation can be brought to the form of a Heun equation. For generic parameters it is a second order differential equation with four regular singular points.
If one then takes the extremal limit $z_2 \rightarrow 0$, the equation becomes a confluent Heun equation, having two regular singularities and one irregular singularity.
However, if at the same time we also take $\omega = m_{\phi} \Omega_{\phi} + \sqrt{z_2} \Delta$,
the equation becomes a hypergeometric equation instead.
In this limit, an asymptotic matching technique first developed in ref. \cite{Lay1999}, and refined in this work, can be employed.
We describe this method in subsection \ref{asymptotic_matching}, using it
to solve the eigenvalue equation to leading order.
In subsection \ref{radial_solution} we apply this method to the radial equation,
expressing the separation constant as a function of the other parameters.

\subsection{The Radial Equation as a Heun Equation}
\label{radial_heun}
A Heun equation is an ordinary differential equation with four regular singular points which are usually taken to be located at $r=0,1,s,\infty$. We can permute these regular singular points using an $SL(2,\mathbb{C})$ transformation.
The Heun equation can always be brought to the canonical form
\small
\begin{align}
\label{heun eq}
&r(r-s)(r-1)F''(r)+ \\
&[\gamma (r-1)(r-s)+\delta  r(r-s)+(\alpha_H +\beta_H -\gamma -\delta +1)(r-1)r]F'(r)+(\alpha_H  \beta_H  r-q)F(r)=0. \nonumber
\end{align}
\normalsize
The condition $\epsilon = \alpha_H + \beta_H - \gamma - \delta + 1$ is needed to ensure regularity of the point at $\infty$.

In the current form of the radial equation (\ref{radial_eq}), it is not evident this
is indeed a Heun equation, therefore we first perform the coordinate transformation $z=x^2$.
We can then rewrite $X$ in terms of $z$ and the parameters $z_2$, $\alpha$, and $a$ as
\begin{align}
X = & \left(z-z_1\right)\left(z-z_2\right) \nonumber \\
z_1= & -\frac{4 a +\left(a^2+1+z_2\right) \alpha }{\alpha }
\end{align}
In terms of the variable $z$, the radial equation indeed has 4 regular singular points, located at $z=0,z_1,z_2, \infty$. The outer horizon of the black hole is located at $z_2$.
The physical region in our equation is thus $z_2 \leq z < \infty$, which we would like to transform to the canonical interval [0,1].
The boundary of AdS is now located at $r=1$ and the black hole horizon is at $r = 0$.
To perform the transformation, we first choose the coordinate transformation
$ r=\frac{z-z_2}{z-z_1}$ under which the singular points transform as:
\begin{equation}
\begin{split}
\infty &\rightarrow 1  \\
z_2 &\rightarrow 0 \\
0 &\rightarrow s \equiv \frac{z_2}{z_1} < 0  \\
z_1 &\rightarrow \infty
\end{split}
\end{equation}
In order to transform the equation to the desired form of (\ref{heun eq}) we also use the ansatz
\begin{equation}
\label{ansatz_radial}
R(r)=(r-1)^{\nu }r^{\mu }\left(r-\frac{z_2}{z_1}\right)^{\rho }F(r).
\end{equation}
Matching to the desired form one finds
that the values of the new parameters are solutions to quadratic equations:
\begin{equation}
\begin{split}
\nu = &\ 0, 2 \\
\mu = &\pm i \frac{-\omega  \left(2 a+\alpha  \left(a^2+z_2\right)\right)+a m_{\phi } \left(2 a +\alpha  \left(1+z_2\right)\right)}{2 \alpha  \sqrt{z_2} \left(-z_1+z_2\right)} \\
\rho= & \pm \frac{a m_{\psi}}{2 \sqrt{z_1} \sqrt{z_2}}.
\end{split}
\end{equation}
Recall that the boundary conditions require the solution be normalizable,
so it will be more convenient to choose $\nu=2$ for the ansatz, which corresponds to $ R(x)\sim \frac{1}{x^4}$ at $\infty$.
In addition, the boundary condition at the horizon requires
the wave be exclusively incoming, i.e that the group velocity be negative.
Therefore we choose
\begin{equation}
\mu = i \frac{- \omega  \left(2 a +\alpha  \left(a^2+z_2\right)\right)+a m_{\phi } \left(2 a +\alpha  \left(1+z_2\right)\right)}{2 \alpha  \sqrt{z_2} \left(-z_1+z_2\right)}.
\end{equation}
Plugging these results into the equation we find the Heun parameters:
\begin{align}
\label{Heun_prmtrs_radial}
s  \equiv& \frac{z_2}{z_1} \nonumber \\
\delta   =&  -1+2\nu = 3 \nonumber \\
\gamma   =&  1+2\mu \nonumber \\
\epsilon   =&  1+2\rho \nonumber \\
\alpha_{H}   =&  \mu +\nu +\rho - i\sqrt{\frac{z_2}{-z_1}}\mu \nonumber \\
\beta_{H}   =&  \mu +\nu +\rho + i\sqrt{\frac{z_2}{-z_1}}\mu \nonumber \\
q  =& \alpha_H \beta_H -\left[\frac{1}{4 \alpha  z_1 \sqrt{z_2}}\left(6 i a  \left(2 l+a \alpha \right)) \omega +6 i a  \alpha  m_{\psi } \sqrt{-z_1}+a^2 \alpha  m_{\phi }^2 \sqrt{z_2}+ \right. \right. \nonumber\\
& \left. \left. \alpha  \left( \lambda + \omega ^2+16 z_1+6 i \omega  \sqrt{z_2}-8 z_2\right) \sqrt{z_2}- \right. \right. \nonumber \\
& \left. \left. 2 i a m_{\phi } \left(3 (2 a+ \alpha )-i  \alpha  \omega  \sqrt{z_2}+3 \alpha  z_2\right)\right)\right].
\end{align}
In the extremal limit, which is of interest to us, we have $z_2\rightarrow 0$, causing the points at $z_2$ and $0$ to merge.
Unfortunately, they do not merge into a regular singular point,
as is evidenced by the fact that all the Heun parameters other than $\delta$
diverge. The equation then becomes a confluent Heun equation.
Little is known of these equations and their solutions.

If we also impose that $\omega = m_{\phi} \Omega_{\phi} + \sqrt{z_2} \Delta$, the situation improves.
The two points now merge into a regular singular point when $z_2 = 0$.
At $z_2 \rightarrow 0$, we have $T\propto\sqrt{z_2}\rightarrow0$.
The restriction to these values of $\omega$ is often used to study near-horizon quasinormal modes,
for example in \cite{Compere2012, Bredberg2010},
and for the solution of a confluent Heun equation,
using matched asymptotic expansion, in \cite{Hod2008,Hod2008a, Hod2009,Hod2010,Hod2011}.

The equation then has 3 regular singular points, and thus can always be transformed into the well-known Hypergeometric equation.
We would like to use this fact to find the quasinormal modes very close to extremality, i.e. for small values of $z_2$.
One can easily check that all Heun parameters have a finite limit when $z_2 \rightarrow 0$.
Indeed, using this scaling ansatz and expanding to zeroth order in $z_2$ one gets
\begin{equation}
\begin{split}
\delta & = 3 \\
\epsilon & =1 \\
\gamma & =1-\frac{i \left(2 a+a^2 \alpha \right) \Delta }{\left(4 a+\alpha +a^2 \alpha \right)} \\
\alpha_{H} & =\frac{1}{2} \left(4+\frac{-i a (2+a \alpha )^2 \Delta +\left(-1+a^2\right) \sqrt{1+a^2+\frac{4 a}{\alpha }} \alpha ^2 m_{\phi }}{(2+a \alpha ) (\alpha +a (4+a \alpha ))}\right)\\
\beta_H & =\frac{1}{2} \left(4-\frac{i a (2+a \alpha )^2 \Delta +\left(-1+a^2\right) \sqrt{1+a^2+\frac{4 a}{\alpha }} \alpha ^2 m_{\phi }}{(2+a \alpha ) (\alpha +a (4+a \alpha ))}\right)\\
q & =\frac{-a (2+a \alpha ) \Delta  (2 i (\alpha +a (4+a \alpha ))+a (2+a \alpha ) \Delta )+\alpha  (\alpha +a (4+a \alpha )) \lambda }{4 (\alpha +a (4+a \alpha ))^2}.
\end{split}
\end{equation}

\subsection{Matched Asymptotic Expansion of the Heun Equation}
\label{asymptotic_matching}
The boundary problem for the Heun equation in the limit of two nearby singularities which merge into a regular singularity, forming a hypergeometric equation, was studied by Lay and Slavyanov in ref. \cite{Lay1999}. In this section we will carry out a general analysis of the Heun equation (\ref{heun eq}) and specialize to our case in the next section.

We begin by following their method and deviate from it at a stage that will be made precise later. The essence of the method is a matched asymptotic expansion using the two regions $r\gg |s|$ and $r \ll 1$ where a solution in the $s\rightarrow 0$ limit can be found.
The small parameter $s$ is used as a scale distinguishing between these two regions but also determines the domain
of mutual validity $|s| \ll r \sim \sqrt{|s|} \ll 1$.

\subsubsection*{The far region $r\gg |s|$}
The far region, $r \gg |s|$, corresponds to spatial infinity in the black hole background.
In this region we can set $s=0$ in (\ref{heun eq}) to obtain the following equation
\begin{equation}
(r-1) r^2 F''[r]+r ((r-1) (1+\alpha_H +\beta_H )+\delta ) F'[r]+(r \alpha_H  \beta_H - q) F[r]=0.
\end{equation}
Here $\alpha_H$,$\beta_H$, etc. refer to the zeroth order contribution, in terms of $s$, of these parameters.
As discussed before, the equation can be brought into a Hypergeometric equation form:
\begin{equation}
\label{hyp eq}
r(1-r)G''[r]+[d-(a+b+1)r]G'[r]-a b G[r]=0.
\end{equation}
Using the ansatz
\begin{equation}
F_{far}[r]=r^{\rho}G[r],
\end{equation}
dividing by $-r^{\rho+1}$ and setting $\rho$ to one of the values:
\footnote{Here and everywhere else in the paper when we use a square root of a complex variable we are working in the following conventions for the choice of the branch cut: $z=r e^{i\theta}$, $-\pi < \theta < \pi$ ; $\sqrt{z}=\sqrt{r}e^{\frac{i\theta}{2}}$. Thus, the real part of the result is always positive.}
\begin{equation}
\label{rho}
\begin{split}
\rho_1= & \frac{1}{2} \left(-\alpha_H -\beta_H +\delta +\sqrt{-4 q_0 +(-\alpha_H -\beta_H +\delta )^2}\right) \\
\rho_2= & \frac{1}{2} \left(-\alpha_H -\beta_H +\delta -\sqrt{-4 q_0 +(-\alpha_H -\beta_H +\delta )^2}\right)
\end{split}
\end{equation}
will bring the equation to the desired form.
For later use, notice that $\rho_1$ and $\rho_2$ satisfy the condition
\begin{equation}
\rho_1+\rho_2=\delta-\alpha_H-\beta_H.
\end{equation}
Choosing $\rho=\rho_1$, which is an arbitrary choice,\footnote{Had we chosen the opposite, $\rho_1$ in the solution for the far region (\ref{solution_far_region}) would be replaced by $\rho_2$. One can check that the resulting solution is the same solution we obtained by choosing $\rho=\rho_1$ written in a different way.} the parameters $a,b,d$ of the hypergeometric equation (\ref{hyp eq}) read
\begin{equation}
\begin{split}
a= & \alpha_H +\rho _1 \\
b= & \beta_H +\rho _1 \\
d= & 1+\alpha_H +\beta_H -\delta +2 \rho _1. \\
\end{split}
\end{equation}
There are two solutions to the Hypergeometric equation (\ref{hyp eq}) around $r=1$.
Since our ansatz (\ref{ansatz_radial}) already picks out the correct behavior near the point $r=1$, we only need to worry about regularity of the solution and thus discard the singular solution at this point. We are left with the solution
\begin{align}
G= & {_2F_1}[a,b,-d+a+b+1;1-r]
\end{align}
where ${_2F_1}$ is the Hypergeometric function.
To sum up, the solution in the far region, to zeroth order in $s$, is
\begin{equation}
\label{solution_far_region}
F_{far}(r)=r^{\rho_1} {_2F_1}[\alpha_H+\rho_1,\beta_H+\rho_1,\delta;1-r]
\end{equation}
where we have used our freedom to normalize the constant coefficient of this solution to $1$ relative to the near region solution.
\subsubsection*{The near region $r \ll 1$ }
The near region, $r \ll 1$, corresponds to the near horizon domain of the black hole background.
In this region, in order to take the limit $s\rightarrow0$ correctly, we need to first make the transformation $r= -s \xi$ in equation (\ref{heun eq}). We then obtain
\begin{align}
&- \xi (- \xi -1)(-s \xi -1)F''[\xi ]-[\gamma (-s \xi -1)(-\xi -1)-\delta  \xi (-s \xi -s)+\nonumber \\
&(\alpha_H +\beta_H -\gamma -\delta +1)(-s \xi -1)(-\xi )]F'[\xi ] +(-\alpha_H  \beta_H  s \xi -q)F[r]=0.
\end{align}
Taking the limit $s\rightarrow0$, the resulting zeroth order equation is
\begin{equation}
-\xi  (1+\xi ) F''[\xi ]-(\gamma +(1+\alpha_H +\beta_H -\delta ) \xi ) F'[\xi ]-q F[\xi ]=0.
\end{equation}
Rewriting the equation in terms of the variable $-\xi$, we find that
it takes the form of a Hypergeometric equation (\ref{hyp eq}) with parameters
\begin{equation}
\begin{split}
a= &-\rho _1 \\
b =& -\rho _2 \\
d= &\gamma.
\end{split}
\end{equation}
Similarly to the far region, one of the solutions of the Hypergeometric equation around zero should be discarded due to singularity at $\xi=0$.
Thus the near region solution is given by
\begin{equation}
F_{near}(\xi)=C F\left(-\rho _1,-\rho _2,\gamma ;-\xi \right).
\end{equation}
\subsubsection*{Matching the two solutions }
The next step is to match the two solutions at the point $r=\sqrt{-s}t$ with $t \sim O(1)$, in the limit $s\rightarrow0$. It is convenient to transform the Hypergeometric functions such their argument approaches zero in the limit $s\rightarrow 0$.

\subsubsection*{Transforming the far region solution}
For the far region we will use the relation, valid when $d-a-b \notin \mathds{Z}$ :
\begin{align}
\label{linear_1}
{_2F_1} (a,b,d,x) = & \frac{\Gamma [a+b-d]\Gamma [d]}{\Gamma [a]\Gamma [b]}(1-r)^{d-a-b} {_2F_1} (d-a,d-b,c-a-b+1,1-x)+ \nonumber \\
& \frac{\Gamma [d-a-b]\Gamma [d]}{\Gamma [d-a]\Gamma [d-b]} {_2F_1} (a,b,-d+a+b+1,1-x).
\end{align}
Using this relation, the far region solution becomes
\begin{align}
\label{zeroth far}
F_{far}(t)=& \left(\sqrt{-s}t\right)^{\rho _2\text{   }}\frac{\Gamma \left[\rho _1-\rho _2\right]\Gamma [\delta ]}{\Gamma \left[\alpha_H +\rho _1\right]\Gamma \left[\beta_H +\rho _1\right]} {_2F_1} \left(\beta_H+\rho_2,\alpha_H+\rho_2,\rho _2-\rho _1+1,\sqrt{-s}t\right)+ \nonumber \\
& \left(\sqrt{-s}t\right)^{\rho _1}\frac{\Gamma \left[\rho _2-\rho _1\right]\Gamma [\delta ]}{\Gamma \left[\beta_H+\rho_2\right]\Gamma \left[\alpha_H+\rho_2\right]}\text{  } {_2F_1} \left(\alpha_H +\rho _1,\beta_H +\rho _1,\rho _1-\rho _2+1,\sqrt{-s}t\right).
\end{align}
\subsubsection*{Transforming the near region solution}
For the near region, taking $r=\sqrt{-s}t$ amounts, in terms of $\xi$, to choosing
$\xi=\frac{t}{\sqrt{-s}}$.
Thus, we will make use of the following formula,
which is valid whenever $a-b \notin \mathds{Z}$, and $\xi \notin (-1,0)$
\begin{align}
\label{linear_2}
{_2F_1} (a,b,d,-\xi ) =& \xi ^{-a}\frac{\Gamma [b-a]\Gamma [d]}{\Gamma [d-a]\Gamma [b]} {_2F_1} \left(a,a-d+1,a-b+1,\frac{-1}{\xi }\right)+ \nonumber \\
& \xi ^{-b}\frac{\Gamma [a-b]\Gamma [d]}{\Gamma [d-b]\Gamma [a]} {_2F_1} \left(b,b-d+1,-a+b+1,\frac{-1}{\xi }\right).
\end{align}
Applying it to the solution we obtained earlier for the near region,
expressed in terms of $t$ and $s$, we get
\small
\begin{align}
\label{zeroth near}
F_{near}(t)=& C \left( \left(\frac{t}{\sqrt{-s}}\right)^{\rho _1\text{   }}\frac{\Gamma \left[\rho _1-\rho _2\right]\Gamma [\gamma ]}{\Gamma \left[\gamma +\rho _1\right]\Gamma \left[-\rho _2\right]} {_2F_1} \left(-\rho _1,-\gamma -\rho _1+1,\rho _2-\rho _1+1,-\frac{\sqrt{-s}}{t}\right)+ \right. \nonumber \\
& \left. \left(\frac{t}{\sqrt{-s}}\right)^{\rho _2\text{  }}\frac{\Gamma \left[\rho _2-\rho _1\right]\Gamma [\gamma ]}{\Gamma \left[\gamma +\rho _2\right]\Gamma \left[-\rho _1\right]} {_2F_1} \left(-\gamma -\rho _2+1,-\rho _2,\rho _1-\rho _2+1,-\frac{\sqrt{-s}}{t}\right)\right).
\end{align}
\normalsize
Now we have everything set to expand  both solutions in powers of $s$ and compare the matching powers of $t$.
We will use the fact that, unless $-d \in \mathds{N}$, ${_2F_1}[a,b,d,s]=1+O(s)$ for $s\rightarrow 0$ to write the two equations resulting from the matching between (\ref{zeroth near}) and (\ref{zeroth far}).  Equating the coefficients of the powers $t^{\rho_1}$ in both equations gives
\begin{equation}
C=\frac{(-s)^{\rho _1}\Gamma \left[\rho _2-\rho _1\right]\Gamma [\delta ]\Gamma \left[\gamma +\rho _1\right]\Gamma \left[-\rho _2\right]}{\Gamma \left[\rho _1-\rho _2\right]\Gamma [\gamma ]\Gamma \left[\delta -\alpha_H -\rho _1\right]\Gamma \left[\delta -\beta_H -\rho _1\right]}.
\end{equation}
\normalsize
Using this result and comparing the powers $t^{\rho_2}$ we obtain (here we deviate from the results in \cite{Lay1999})
\small
\begin{equation}
\frac{\left(\Gamma \left[\rho _1-\rho _2\right]\right){}^2\Gamma \left[\gamma +\rho _2\right]\Gamma \left[-\rho _1\right]\Gamma \left[\beta_H+ \rho_2 \right]\Gamma \left[\alpha_H + \rho_2\right]}{\Gamma \left[\alpha_H +\rho _1\right]\Gamma \left[\beta_H +\rho _1\right]\left(\Gamma \left[\rho _2-\rho _1\right]\right){}^2\Gamma \left[\gamma +\rho _1\right]\Gamma \left[-\rho _2\right]}=\left(\sqrt{-s}\right)^{2\rho _1-2\rho _2}.
\label{zeroth}
\end{equation}
\normalsize
Assuming $\Re (\rho_1-\rho_2 )> 0$, the RHS of (\ref{zeroth}) goes to zero as $s\rightarrow 0$, so the denominator of the LHS must diverge in when taking the limit. This means one of the arguments of the gamma functions in the denominator should be equal to a negative integer which gives a quantization condition.\\

Up until now the discussion was completely general.
In the next section we will focus on the radial equation,
expressing the quantization condition explicitly.

\subsection{Applying the Asymptotic Matching to the Radial Equation}
\label{radial_solution}
In order to extract the correct quantization condition for $\lambda$ from equation
(\ref{zeroth}) we must check several cases, some of which do not lead to a valid solution.
As we have shown, the matching procedure implies that one of the following should be integer:
\begin{align}
\beta_H+\rho_{1} & =\frac{1}{2} \left(3+\sqrt{1-\frac{\alpha  \lambda }{\alpha +a (4+a \alpha )}}-\frac{\left(-1+a^2\right) \sqrt{1+a^2+\frac{4 a}{\alpha }}\text{  }\alpha ^2 m_{\phi }}{(2+a \alpha ) (\alpha +a (4+a \alpha ))}\right) \label{radial_quant_4} \\
\alpha_H+\rho_1 & =\frac{1}{2} \left(3+\sqrt{1-\frac{\alpha  \lambda }{\alpha +a (4+a \alpha )}}+\frac{\left(-1+a^2\right) \sqrt{1+a^2+\frac{4 a}{\alpha }}\text{  }\alpha ^2 m_{\phi }}{(2+a \alpha ) (\alpha +a (4+a \alpha ))}\right) \label{radial_quant_1}\\
\gamma+\rho_1 & =\frac{1}{2} \left(1-\frac{i a (2+a \alpha ) \Delta }{\alpha +a (4+a \alpha )}+\sqrt{1-\frac{\alpha  \lambda }{\alpha +a (4+a \alpha )}}\right) \label{radial_quant_2}\\
-\rho_2 & =\frac{1}{2} \left(1-\frac{i a (2+a \alpha ) \Delta }{\alpha +a (4+a \alpha )}+\sqrt{1-\frac{\alpha  \lambda }{\alpha +a (4+a \alpha )}}\right) \label{radial_quant_5}\\
\rho_2-\rho_1 & =-\sqrt{1-\frac{\alpha  \lambda }{\alpha +a (4+a \alpha )}} \label{radial_quant_3}.
\end{align}
The expressions possibly consistent with this condition are\footnote{Since $a<1$, the expression in (\ref{radial_quant_4}) is manifestly positive, also it is easy to see that $\gamma+\rho_1=-\rho_2$, so both expression (\ref{radial_quant_2}) and expression (\ref{radial_quant_5}) will provide the same quantization condition.} (\ref{radial_quant_1}), (\ref{radial_quant_2}) and (\ref{radial_quant_3}).
In Appendix \ref{Appendix_A} we show the expression (\ref{radial_quant_3}) doesn't lead to a valid quantization condition.

\subsubsection*{Quantization condition for small $m_\phi$}
We are interested in rapidly rotating black holes, i.e\ $1-a^2<<1$. Thus for finite fixed $m_{\phi}$ the third term in (\ref{radial_quant_1}), which is the only negative term with a negative real part in the expression, is very small compared to the other terms making (\ref{radial_quant_1}) positive. Therefore, the only possibility for small $m_{\phi}$ is $\gamma+\rho_1=-n$ which gives the following quantization condition for $\lambda$:
\begin{equation}
\label{radial_sol}
\lambda=\frac{-4 n (1+n) (\alpha +a (4+a \alpha ))+2 i a (1+2 n) (2+a \alpha ) \Delta }{\alpha }+\frac{a^2 (2+a \alpha )^2 \Delta ^2}{\alpha (\alpha +a (4+a \alpha ))},
\end{equation}
or, more compactly
\begin{equation}
\label{radial_sol1}
\Delta=\frac{\alpha+4a+a^2 \alpha}{2a+a^2\alpha} \left(\pm \sqrt{\frac{\alpha \lambda}{\alpha+4a+a^2 \alpha}
-1} -i (1+2n)\right).
\end{equation}
Taking into account $\Re \left(\rho_1- \rho_2\right) >0$, which is necessary for the validity of the calculation, we get the additional restriction
\begin{equation}
\begin{split}
\Re( -n-\gamma -\frac{1}{2}\left(-\alpha_H-\beta_H +\delta \right)) = \frac{1}{2}\Re\left(-1-2 n+\frac{i a (2+a \alpha ) \Delta }{\alpha +a (4+a \alpha )}\right) > 0.
\end{split}
\end{equation}
This in turn implies
\begin{equation}
\label{restriction_1}
\Im\Delta \sqrt{z_2}  < -\frac{4a+\alpha+a^2 \alpha}{2a+a^2 \alpha} \sqrt{z_2} \left(2n+1\right)= -2\pi T \left(2n+1\right)
\end{equation}
where $T$ is the temperature of the black hole.
In particular, this means $\Im \Delta <0$, so the chosen quantization condition gives only stable solutions.

Thus we obtain that $\Delta$ is always purely imaginary which implies
\begin{equation}
\Re \omega = m_{\phi} \Omega_{\phi}
\end{equation}
for any parameters and small enough $m_{\phi}$.
Finally, using the restriction (\ref{restriction_1}), we can eliminate the solution with a plus sign in front of the square root in \ref{radial_sol1}.

\subsubsection*{Quantization condition for large $m_\phi$}
Turning to $m_{\phi}>\frac{1}{1-a^2}$, taking $\alpha_H+\rho_1=-n$
we obtain an expression for $\lambda$ which is independent of
$\Delta$.  This means that at leading order the angular equation won't include $\Delta$, and one needs to compute the next-to-leading order in $\sqrt{z_2}$ to extract
a quantization condition for it.

As in the previous case we apply the condition $\Re \left(\rho_1- \rho_2 \right) >0$ to get
\begin{equation}
-n-\alpha_H-\frac{1}{2}\left(-\alpha_H-\beta_H +\delta \right)= -\frac{3}{2}-n-\frac{a m_{\phi }}{\sqrt{1+a^2+\frac{4 a}{\alpha }}}+\frac{\sqrt{1+a^2+\frac{4 a}{\alpha }} \alpha  m_{\phi }}{4+2 a \alpha }>0
\end{equation}
and the restriction on $m_{\phi}$, coming from the demand that the solution be compatible with $n\geq0$, is
\begin{equation}
\label{m_phi_con}
m_{\phi}> \frac{3 (2+a \alpha ) \sqrt{\alpha +a (4+a \alpha )}}{\left(1-a^2\right) \alpha ^{3/2}}.
\end{equation}
This is the quantitative interpretation of the requirement for $m_{\phi}$ to be large in order to ensure the validity of the quantization condition. $D$ is a positive quantity, and so no instability is caused by the quasinormal modes obtained in this limit.

To sum up, there are two valid quantization conditions one of which always gives stable solutions and is valid for any $m_{\phi}$.
The other condition is valid only for large enough $m_{\phi}$s and one must employ the next to leading order to find the QNMs.
Fortunately, in the limit of interest, i.e. of rapid rotation $a\rightarrow 1$, equation (\ref{m_phi_con}) implies that these modes are pushed to infinite $m_{\phi}$ for finite small enough $z_2$, and they disappear from the spectrum. We will therefore ignore them in this paper, which analyzes low energy excitations of the black hole, and we will revisit them in a later paper \cite{work_in_progress}, which analyzes high frequency modes in a slightly different limit using numerical tools .

\section{The Angular Equation}
\label{angular_eq_sec}
The angular equation is given by
\begin{align}
\label{angular_eq}
\frac{1}{ y }\partial _y\left(y Y \partial_yS(y)\right) - \left( \frac{\left(y^2 \left(a m_{\phi }- \omega \right)-am_{\phi }+a^2\omega \right)^2}{Y}+\frac{a^2 m_{\psi}^2}{ y^2 }+\lambda \right) S(y)=0.
\end{align}
As we discuss below, it can also be brought to the canonical form of a Heun equation, shown in (\ref{heun eq}).
In subsection \ref{angular_heun} we show the explicit transformation.
However, there does not appear to be a limit, consistent with the $z_2\rightarrow 0$ limit we took in the radial equation, in which it simplifies and allows for a general analytic solution.
We therefore use the Continued Fraction Method (CFM), described in subsection \ref{continued_fraction}, to solve the eigenvalue problem and
find the quantization condition. In subsection \ref{angular solution} we apply the method to the angular equation,
and obtain numerical solutions for eigenvalues, as well a an analytic result in the case $\alpha = 2$.

\subsection{The Angular Equation as a Heun Equation}
\label{angular_heun}
In terms of the variable $z=y^2$ the equation is also a linear ODE with four regular singular points: $z = 0, a^2, 1, \infty $.
The physical region of interest is $ 0\leq z \leq a^2$, which we would like to transform it to the interval $[0,1]$.
Thus, we use the transformation $r=\frac{z}{a^2}$ under which the singular points transform as:
\begin{align}
& 0 \rightarrow 0 \nonumber \\
& a^2 \rightarrow 1 \nonumber \\
& 1 \rightarrow \frac{1}{a^2} >1. \\
\end{align}

Next, we plug in the ansatz
\begin{equation}
\label{angular_ansatz}
S(r)=r^{\kappa }(r-1)^{\eta}\left(r-\frac{1}{a^2}\right)^{\chi }H(r)
\end{equation}
and using any of the following choices
\begin{align}
\eta & = \pm \frac{m_{\phi}}{2} & \kappa & = \pm \frac{m_{\psi}}{2} & \chi & = \pm \frac{\omega }{2}
\end{align}
the equation takes the form (\ref{heun eq}).
In order to have regularity in the interval $y \in [0,a]$ the preferred choice in our case is $\eta = \frac{|m_{\phi}|}{2}$ and $\kappa = \frac{|m_{\psi}|}{2}$.
The parameters of the Heun equation, as defined in equation (\ref{heun eq})
\begin{equation}
\label{angular_heun_eq}
\begin{split}
s= & \frac{1}{a^2} \\
\alpha_{H} = & \kappa +\eta +\chi  =  \frac{1}{2} \left( |m_{\phi }|+|m_{\psi }|- \omega \right)  \\
\beta_{H} = & 2 + \kappa +\eta +\chi  =  \frac{1}{2} \left(4+|m_{\phi }|+|m_{\psi }|- \omega \right)  \\
\delta = & 1+2\eta  =  1+|m_{\phi }|  \\
\gamma = & 1+2\kappa  =  1+|m_{\psi }|   \\
\epsilon = & 1+2\chi =1- \omega   \\
q = & \frac{-2 a^2 \omega -2  \left(\left(-1+a^2 \omega \right) \left|m_{\psi }\right|- |m_{\phi }| \left(1+\left|m_{\psi }\right|\right)\right)}{4 a^2}+  \\
& \frac{1}{4 a^2}\left(2 \left|m_{\psi }\right|+m_{\psi }^2\right)+\frac{1}{4a^2}
\left(\lambda +\left(-a \omega +m_{\phi }\right)^2+m_{\psi }^2\right).
\end{split}
\end{equation}
Both choices of $\chi$ should produce the same results.
It is more convenient however to choose $\chi= -\frac{\omega}{2}$ in order to maintain the explicit
dependence in the parameters on $\omega - m_{\phi}$ for $m_{\phi} > 0$.\footnote{The alternative choice can be reproduced by reinstating $l$, as appropriate by dimensional analysis, and making the transformation to $l \rightarrow -l$ in these equations.}

\subsection{The Continued Fraction Method for the Heun Equation}
\label{continued_fraction}
There are two local solutions to Heun's equation (\ref{heun eq}) in the neighborhood of
the singular point $r=0$ and they take the following form\footnote{The same is true for the point $r=1$ with the replacement of $\gamma$ by $\delta$ and of $r$ by $1-r$.}
\begin{equation}
\begin{split}
H^0_1(r)= & \sum\limits_{k=0}^{\infty} a_k r^{k} \\
H^0_2(r)= & r^{1-\gamma}\sum\limits_{k=0}^{\infty} b_k r^{k}.
\end{split}
\end{equation}
The relevant boundary conditions for our problem is that the function approaches a constant at $r=0$ and $r=1$, as we have stripped off the angular dependence in the ansatz (\ref{angular_ansatz}).
Recall that $s>1$ in our case.
Plugging the series solution around $r=0$, denoted previously by  $H_1^0(r)$, back into the Heun equation results in a 3-term recurrence relation:
\begin{equation}
\label{recurrence}
f_k a_{k+1}+g_k a_k+h_k a_{k-1}=0,
\end{equation}
with the recursion coefficients given by
\begin{equation}
\begin{split}
\label{rec_coef}
f_k = &s (k+1)(k+\gamma)  \\
g_k = &-k \left[s \left(k+\gamma+\delta-1\right)+k+\gamma+\epsilon-1\right]-q \\
h_k = & (k+\alpha_H-1)(k+\beta_H-1).
\end{split}
\end{equation}

The initial conditions for this series are
\begin{equation}
\begin{split}
\label{initial_con}
& a_{-1}=  0 \\
& a_{0}=  1.
\end{split}
\end{equation}
Such three term recurrence relations, in the context of solutions to differential equations, were studied thoroughly by
Leaver \cite{Leaver85, Leaver86} and we will follow his methods to obtain a numerical solution.

Within this method, regularity at $r=1$ restricts the coefficients in (\ref{rec_coef})
which in turn determines the quantization of $\lambda$ via the dependence of $q$ on $\lambda$.

\subsubsection*{Regularity at the point $r=1$}
Indeed, for the series solution to be a genuine solution it must also satisfy the boundary conditions, and in particular be regular at $r=1$. This implies the convergence of $\sum\limits_{k=0}^{\infty}a_k $.
To check whether the series converges we examine the asymptotic behavior of  $|r_k|\equiv|\frac{a_{k+1}}{a_{k}}|$.
Denoting
\begin{align}
t= &\lim\limits_{k\rightarrow\infty}r_k = \lim\limits_{k\rightarrow\infty}r_{k+1} \nonumber \\
a = &\lim\limits_{k\rightarrow\infty}\frac{g_k}{f_k} \nonumber \\
b= &\lim\limits_{k\rightarrow\infty}\frac{h_k}{f_k}
\end{align}
we rewrite equation (\ref{recurrence}) in the form:
\begin{equation}
t^2+a t +b=0.
\end{equation}
The two solutions to this equation give us two types of asymptotic behaviors for $r_k$.

Actually, any three term recurrence relation has two independent solution sequences (without stating the initial conditions) $A_k, B_k$. These frequently have the property that $\lim\limits_{k\rightarrow \infty} \frac{A_k}{B_k}=0$ in which case $A_k$ is called the minimal solution. If $|t_1|>|t_2|$ then the solution with the asymptotic behavior $t_2$ is the minimal one. There is a unique choice of initial conditions corresponding to the minimal solution while all other choices inevitably produce the dominant one.
As we explain below, the correct solution to the problem is the minimal one. Therefore, since the initial conditions are set and are independent of any parameter, as seen in (\ref{initial_con}), we will need to impose a restriction on the recurrence coefficients (\ref{rec_coef}) such that the resulting series solution is minimal given these initial conditions. This will provide the required quantization condition for $\lambda$.

Applying the analysis above to the Heun equation we have
\begin{equation}
\begin{split}
\label{t}
t_1 = & 1 \\
t_2= & \frac{1}{s}.
\end{split}
\end{equation}
Recalling our convention, $s>1$  we see that the minimal solution will always give a convergent series at $r=1$.

However, the dominant solution might also give a regular solution at $r=1$.
To check the convergence of the dominant solution we retain the $\frac{1}{n}$ terms in $a$ and $b$ to get
\begin{equation}
t_1=1 + \frac{-2+\delta}{n}
\end{equation}
Thus, depending on the value of $\delta$, the dominant solution might give a convergent series at $r=1$. However, regularity alone is not always enough\footnote{In the example of the angular equation we will treat below, regularity is enough to uniquely choose the minimal solution. However we would like to show that the minimal solution is the unique solution also in the more general setting.} and, as described in \cite{Starinets:2002br}, using the full boundary condition at $r=1$ it can be proven that the minimal solution is the correct solution of the problem for any $\delta$.

\subsubsection*{Taking into account the boundary condition at $r=1$}
For completeness we repeat the argument of \cite{Starinets:2002br} here. Returning to the boundary condition at $r=1$ we recall we have assumed that the corresponding power law behavior was stripped off, so for $r=1$ the function should simply go to one. To find the restrictions stemming from this demand we look at the two local solutions around the point $r=1$, they have the properties $H^1_1(r)\sim 1+ \cdot \cdot \cdot$ and $H^1_2(r)\sim (1-r)^{1-\delta}+\cdot \cdot \cdot$.
Any differential equation of second order has only two linearly independent solutions. In the interval $[1,s]$ the local solution around zero, which we considered thus far, should be a linear combination of the two local solutions around $r=1$. Moreover, to satisfy the boundary condition it must be exactly equal to $H^1_1(r)$, which implies it should be absolutely convergent up to $r=s$ and not only up to $r=1$ (and also that it is equal to what is called in the literature a Heun function).
Examining equations (\ref{t}) it is clear this requirement is satisfied only for the minimal solution. Thus, we have shown that for a perturbation of a black hole described by the Heun equation, in the context of the CFM method, the correct solution in the whole range of the parameters is the minimal one.

\subsubsection*{Obtaining the quantization condition}
Returning to the issue of a quantization condition, we quote a known result for three term recurrence relations.
The following equation holds for the minimal solution of the recursion relation  (\ref{recurrence}),
and in particular the continued fraction shown below is convergent \cite{Gautschi1967}
\begin{equation}
\label{cont_frac}
\frac{a_{k+1}}{a_{k}}=\cfrac{-h_{k+1}}{g_{k+1}-\cfrac{f_{k+1}\cdot h_{k+2}}{g_{k+2}-\cfrac{f_{k+2}\cdot h_{k+3}}{g_{k+3}-...}}}\equiv
-\frac{h_{k+1}}{g_{k+1}-}\cdot\frac{f_{k+1} h_{k+2}}{g_{k+2}-}\cdot\frac{f_{k+2} h_{k+3}}{g_{k+3}-}...
\end{equation}
Using this equation for $k=0$ together with the equation $\frac{a_1}{a_0}=-\frac{g_0}{f_0}$ (arising from the initial conditions (\ref{initial_con}) and the recurrence relation) we get the condition:
\begin{equation}
\label{quant_numer}
0=g_0-\frac{f_0\cdot h_1}{g_1-} \cdot\frac{f_1\cdot h_2}{g_2-} \cdot\frac{f_2\cdot h_3}{g_3-}...
\end{equation}
Truncating the continued fraction at some $k$ we thus get a polynomial equation for $\lambda$ which can be solved numerically.
Since the continued fraction is convergent, some of the $\lambda$s found by solving this equation correspond to true eigenvalues,
with the smallest ones found first.
Increasing the cutoff causes more QNMs to be found, and also results in better accuracy for the ones already found.

\subsection{Applying the Continued Fraction Method to the Angular Equation}
\label{angular solution}
Before using the CFM to solve the angular equation we first need to take the same limit we took when dealing with the radial equation. In particular, we replace $\omega=m_{\phi} \Omega_{\phi}+\sqrt{z_2}\Delta$ and $m_{\psi}=0$ into the relations (\ref{angular_heun_eq}) and expand to zeroth order in $z_2$.
The parameter $\Delta$ enters these relations only through $\lambda$ in this order.
Thus, one can proceed to solve (\ref{quant_numer}) numerically for $\lambda$, getting quantized values for it. Then, rewriting  (\ref{radial_sol1}) to express $\Delta$ in terms of $\lambda$, one can use each of the values found for $\lambda$ to extract another quantized spectrum for $\Delta$.

When carrying out this procedure, there is one special case.
Setting either $\alpha=2$ or $a=1$, after taking the limit discussed above, implies
$\Omega_{\phi} = 1 + O(z_2)$. This in turn means that  $\omega=m_{\phi}$ to zeroth order in $z_2$ and one obtains in (\ref{angular_heun_eq})
\begin{equation}
\begin{split}
&\alpha_H=0 \\
&\beta_H=2.
\end{split}
\end{equation}
Note that the above relations hold also if one sets $m_{\phi}=0$ or takes $a \rightarrow 1$.
In the recurrence relation (\ref{recurrence}), $h_k$ now takes the form
\begin{equation}
h_k=(-1 + k) (1 + k).
\end{equation}
In particular, we have $h_1=0$ .

Plugging $h_1=0$ into the recurrence relation (\ref{recurrence}) we find that the equation (\ref{cont_frac}) is no longer valid for $k=0$. Instead one obtains an initial condition for $k=1$, which is
\begin{equation}
\frac{a_2}{a_1}= \frac{-g_1}{f_1}
\end{equation}
This condition can be used to derive a quantization condition
in the same fashion it was derived for $k=0$. The new quantization condition is
\begin{equation}
0=g_1-\frac{f_1\cdot h_2}{g_2-} \cdot\frac{f_2\cdot h_3}{g_3-}...
\end{equation}

However, a much more interesting consequence of the relation $h_1 = 0$ is that, if we assume
\begin{equation}
\label{g0=0}
g_0=0
\end{equation}
and use the recurrence relation (\ref{recurrence}) at $k=0$ with the initial condition $a_{-1}=0$, we get that $a_1=0$. Then, for $k=2$, the recurrence relation implies that $a_2=0$ which means that all coefficients of the series are zero except for $a_0$ which is equal to 1. Thus, the assumption (\ref{g0=0}) leads to the solution
\begin{equation}
H(r)=1
\end{equation}
which meets the requirements of the boundary conditions and thus is a legitimate and exact solution of the problem.
Moreover, (\ref{g0=0}) supplies the corresponding equation for the eigenvalue $\lambda$, which can be solved analytically.

A simpler way to see $H(r)=1$ solves the equation is to examine (\ref{heun eq}) under the condition $\alpha_H=0$. It is evident that a constant solves this equation if both $q=0$ and the boundary conditions are $H(0)=H(1)=const$. Indeed, the condition $q=0$ is equivalent to (\ref{g0=0}) and in our case the boundary conditions are $H(0)=H(1)=1$. As mentioned above, this also occurs when $a\rightarrow1$ or $m_{\phi}=0$.

Indeed, we obtain the value of the separation constant
\begin{equation}
\lambda= -m_{\phi}(1-a)[2(1+a)+m_{\phi}(1-a)].
\end{equation}
Plugging $\lambda$ into equation (\ref{radial_sol1})\footnote{As we discuss in the paragraph below (\ref{radial_sol1}), the correct choice of sign in (\ref{radial_sol1}) is a minus.} we find
\begin{equation}
\Delta=i\frac{-2 (1+a) (1+n)+(-1+a) m_{\phi }}{a}
\end{equation}

\section{Results}
\label{results_sec}
We have analyzed the equation of motion of a massless uncharged scalar in the background of a charged rotating black hole. The equation of motion for such a scalar separates into an angular and a radial part, and we have denoted the separation constant by $\lambda$.
The radial equation can be solved in the extremal limit, $z_2 \rightarrow 0$,
along with $\omega = m_{\phi} \Omega_{\phi}+\sqrt{z_2} \Delta$ and $m_{\psi}=0$.
As noted in \cite{Birkandan2011} for similar black holes, in the above limit the angular equation does not depend on $\Delta$.
The eigenvalue problem for $\lambda$ is thus decoupled, and one can solve it
independently, subsequently
using the result to solve the eigenvalue problem for $\Delta$ in the radial equation.

\subsubsection*{The Radial Equation}
Solving this equation using the matched asymptotic expansion
yields the following relation between $\Delta$, and hence $\omega$, and $\lambda$
\begin{equation}
\label{expression_omega}
\omega= m_{\phi} \Omega_{\phi}-i 4\pi T \left(\frac{1}{2}\sqrt{1-\frac{\alpha \lambda(\alpha,a,m_{\phi})}{\alpha+4a+a^2 \alpha}} +\frac{1}{2}+n\right) \equiv m_{\phi} \Omega_{\phi}-
i 4\pi T\left(D\left(a,\alpha, m_{\phi}\right)+n\right)
\end{equation}
expressed in terms of the temperature $T$ of the black hole, which is used here as an independent variable instead of $z_2$. In this limit $T=\sqrt{z_2}(a+1)/(2\pi a)$.
The deviation of $\omega$ from $m_{\phi} \Omega_{\phi}$ must be purely imaginary in order to ensure validity of the solution. Apart from its dependence on $m_{\phi}$, this is the same spectrum as for an operator of dimension $D$ in a finite temperature chiral $(1+1)$ CFT.

\subsubsection*{The Angular Equation}
The angular equation can be used to solve for $\lambda$ in terms of $m_{\phi}$ and $a$.
If one assumes either  $\alpha=2$, $a\rightarrow1$ or $m_{\phi}=0$
it is possible to obtain some eigenvalues and eigenfunctions analytically,
while for a generic choice of these parameters we must
calculate $\lambda$ only numerically.

\subsubsection*{Numerical Results for the Quasinormal Modes}
\begin{figure}[H]
    \includegraphics[width=1 \linewidth,scale=1]{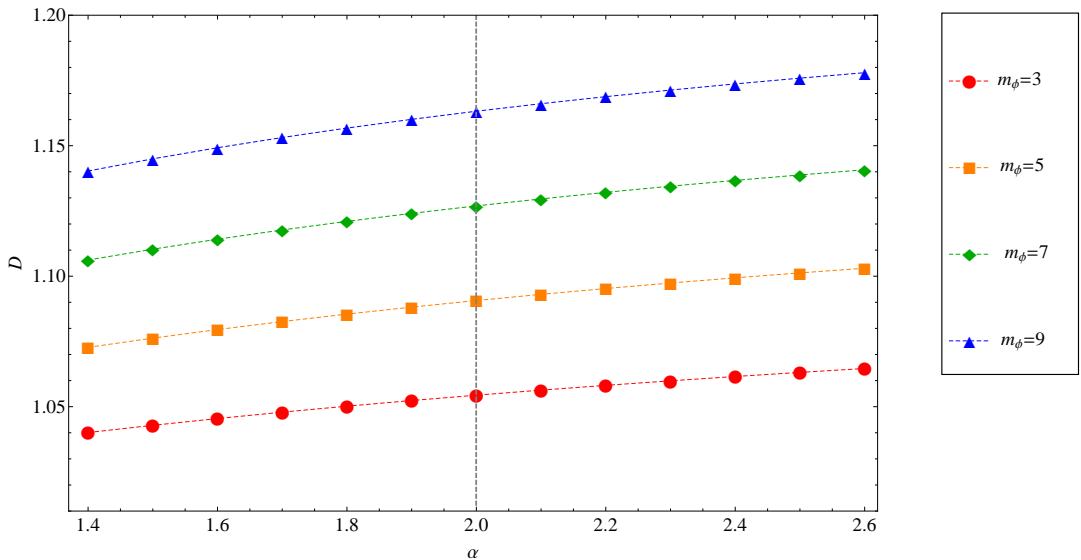}
    \centering
    \caption{The parameter D for the first quasinormal mode with $n=0$  and $a=0.93$. The grey line marks $\alpha=2$ for which we obtained analytical results as well. }
    \label{qnm1}
\end{figure}

We now list some numerical results obtained using the continued fraction method.
The computations have been performed using {\it Mathematica}, which allowed for high precision numerics.
The parameters are chosen to be $\alpha = 2$, $m_{\phi} = 5$ and $a = 0.93$.
The choice of $\alpha$ allows us to compare the lowest mode with analytical results.
The results displayed are for $D$, with $\omega = m_{\phi} -4\pi i T(D + n)$, and they are valid for $T << 1-a^2$.
\begin{table}[H]
	\centering
		\begin{tabular}{||c||c|c|c|c|c||}
			\hline
            $\lambda$ & $-1.47$ & $-11.17$ & $-24.86$ & $-42.32$ & $-63.35$ \\
            \hline
			$D$ & $1.09$ & $1.50$ & $1.89$ & $2.26$ & $2.62$ \\
			\hline
		\end{tabular}	
    \caption{Computed values of the separation constant $\lambda$ and the offset $D$, computed using the CFM for $\alpha = 2$, $m_{\phi} = 5$ and $a = 0.93$.}
\end{table}
In Figure \ref{qnm1} we plot the first quasinormal mode for $n=0$ and $a=0.93$ for several $\alpha$s and $m_{\phi}$s.

As was previously mentioned, the singularity remains at a finite distance even when the extremal limit is taken, when one approaches the horizon radially along $y=0$.
One may expect that in such a case there would be no
decoupled near-horizon limit, which occurs when the horizon is pushed to infinite proper distance.
Performing the analysis we find that as $z_2$ tends to zero the number of modes grows, and correspondingly they become denser. This is easy to see because the spectrum of $\lambda$, and therefore $\Delta$ is independent of $z_2$ in this limit, and as $z_2 \rightarrow 0$ it condenses around $m_{\phi} \Omega_{\phi}$.
This is an indication that modes on this background have a decoupling limit, similar to that of an $AdS$ throat.

\subsubsection*{Analytic Results for $\alpha=2$}
\begin{itemize}
\item
For the case of $\alpha=2$ we obtained an analytic expression for one of the eigenvalues $\lambda$,
given by
\begin{equation}
\lambda= -m_{\phi}(1-a)[2(1+a)+m_{\phi}(1-a)].
\end{equation}
The analytic expression for the eigenfrequencies in this family of modes is given by
\begin{equation}
\begin{split}
\label{lowest_omega}
\omega=& m_{\phi}+i\sqrt{z_2} \left(-\left(2+2n\right)\left(1+\frac{1}{a}\right)+m_{\phi}\left(1-\frac{1}{a}\right)\right) \\
& = m_{\phi}-i 4 \pi T \left(1+\left(\frac{m_{\phi }(1-a)}{2(1+a)}\right) + n\right).
\end{split}
\end{equation}
\item Using the numerical procedure, we checked several sets of parameters in which the analytic solution exists. Sorted by their absolute value, the mode obtained analytically is the lowest of the bunch. Thus, in terms of $\omega$, we get a family of modes parameterized by $n$, with $n=0$ corresponding to the mode with slowest decay.

\item The analytic expression for the angular part
of the eigenfunction of the lowest stable mode is given by
\begin{equation}
\label{angular_func}
S(y)=\left(y^2-a^2\right)^{\frac{m_{\phi}}{2}}\left(y^2-1\right)^{-\frac{m_{\phi}}{2}}=
(a^2sin^2\theta)^{\frac{m_{\phi}}{2}}(1-a^2cos^2\theta)^{\frac{-m_{\phi}}{2}}.
\end{equation}
where the change of variables from $y$ to $\theta$ is $y^2=a^2 cos^\theta$.
\end{itemize}

\subsubsection*{Results in the Large Charge Limit - $a\rightarrow 1$}
As previously mentioned, one can also obtain some analytical results
upon taking the additional limit $a \rightarrow 1$ or setting $m_{\phi}=0$.
It is easily seen that in this case, $\lambda=0$ is a valid eigenvalue with $S(y) = const$,
and consequently, for this mode $D=1$.

In fact, when $a\rightarrow 1$, the angular equation does not depend on $m_{\phi}$ or $\alpha$.
The dependence on $\alpha$ enters only through $\Omega_{\phi}$, which approaches $1$ in this limit,
and $m_{\phi}$ then enters the equation solely through the combination $(1-a)m_{\phi}$.
Thus, we obtain a universal spectrum, to leading order in $1-a$, given by
\begin{equation}
\omega=m_{\phi}-i4\pi T \left(D(\alpha)+n\right),
\end{equation}
with the dependence of $D$ on $\alpha$ obtained using (\ref{expression_omega})
\begin{equation}
D(\alpha) = \frac{1}{2}\sqrt{1-\frac{\alpha \lambda_{u} }{2\alpha+4}} +\frac{1}{2}.
\end{equation}
Here, $\lambda_{u}$ is the universal angular spectrum in the $a \rightarrow 1$ limit.
We have computed its first lowest-lying values, which are given in the following table.
\begin{table}[H]
	\centering
		\begin{tabular}{||c||c|c|c|c|c||}
			\hline
            $\lambda_u$ & $0$ & $-10.12$ & $-25.44$ & $-44.95$ & $-68.17$ \\
			\hline
		\end{tabular}	
    \caption{Numerical values for the lowest modes in the universal spectrum $\lambda_u$.}
    \end{table}

\subsubsection*{Some General Features of the Radial Part of the Wave Functions}
An expression for the radial part of the wave function, as a function of the separation constant,
was obtained in Section \ref{asymptotic_matching} for two different regions of the $r$
segment and a matching between them was performed.
The exact expressions for the function in the two
different regions when $\alpha=2$, and an analytic expression is available,
are written in Appendix \ref{Appendix_B} ,
here we only describe the most prominent features of these solutions.

First, we consider the region stretching from infinity to a distance of $-s=-\frac{z_2}{z_1}$ from the horizon. In terms of $r=\frac{x^2}{x^2+(1+a)^2}$, this is the region $\sqrt{-s} < r < 1$.
We find that in this region the function is a polynomial in $r$ and is independent of $n$.
Thus, far from the horizon the wave function $\Phi(r,\theta)=e^{-i(\omega \tau-m_{\phi} \phi)}R(r)S(\theta)$ depends on $n$ only through its time dependence.

It is important to mention this is true not only for the lowest
eigenvalue for $\lambda$, which we found analytically,
but also for the rest of the modes.
This is a consequence of the fact that $\lambda$ is independent of $n$, thus if one changes $\lambda$
it cannot affect the $n$ dependence of the eigenfunction.
Furthermore, this statement can be verified also for
$\alpha\neq 2$ since the $n$ dependence can be determined from the expression for the radial function without specifying $\lambda$. Indeed, we find it is independent of $n$. The precise expression in terms of $\lambda$ is given in Appendix \ref{Appendix_B}.
Thus, the eigenfunctions away from the horizon are all independent of $n$, and couple to all the near-horizon modes.

In the near horizon region, $0 < r < \sqrt{-s} $, the most prominent feature of the solution is that when one approaches the horizon, $r = 0$, it takes the form $R(r) \sim \frac{r^{-D-n}}{z_2^{-n}}+ O(\frac{r}{z_2})$ where $D$ is defined in \ref{expression_omega}. One implication of this result is that the solution diverges for $n>0$ as one gets closer to the horizon. This is a general statement for decaying quasinormal modes in the background of a black hole. One way to see this is to examine the behavior near the horizon.
For a scalar field one usually gets
\begin{align}
\label{phi_near_horizon}
& \Phi \sim r ^{\pm A( i\omega +B)} e^{-i\omega \tau}  & r\sim 0
\end{align}
where A and B are independent of $\omega$ and $A>0$. The relation is written using coordinates in which the horizon is at $r=0$. The boundary conditions at the horizon require a purely ingoing wave which leaves us with the minus sign in (\ref{phi_near_horizon}) (cf. \cite{Cardoso2004}) . Thus for a decaying mode, i.e $\Im \omega <0$, since $\omega$ is the only complex quantity in the equation, $r$ appears with a negative power in $\Phi$ near the horizon. Indeed this phenomena appears in other works on quasinormal modes of black holes, such as \cite{Leaver85} and \cite{Cardoso2004}.

\section{Summary and Discussion}
\label{discussion_sec}
This work analyzes the equation of motion for a massless minimally coupled scalar in the background of a charged rapidly rotating black hole in the limit of $T \rightarrow 0$, which corresponds to an extremal zero-entropy degeneration of a black hole.

We began by examining Fermi surface operators in the $PSU(1,1|2)$ sector
of ${\cal N}=4$ $SU(N)$ SYM theory.
These operators have been constructed explicitly out of the fermionic operators of this sector and, due to its extended symmetry, are eigenstates of the dilatation operator.
They are also known to have, in the large charge limit, small corrections to the conformal dimension.
One may hope that this would allow matching them to gravity configurations of similar properties.

A conjecture has then been made regarding the gravity solutions dual to these black holes.
By studying black holes which are solutions of a consistent truncation of type $IIB$ SUGRA on $S^5$,
the dual configurations can be identified using scaling laws between the angular momentum and
charge of the Fermi surface operators.
In order to verify the correspondence, we compute the quasinormal modes of an massless scalar field.
For the result to have a meaning in the dual CFT, the inclusion of the scalar must also reside
in a consistent truncation of the full theory. The addition dilaton (or axion) forms such a consistent truncation.

The limit $T \rightarrow 0$ corresponds to a singular background, as the ring of singularities, located along one of the independent rotation planes, reaches the horizon. The area of the horizon also shrinks to zero in that limit, forming a zero entropy degeneration. Nonetheless, it is possible to solve the equations of motion semi-analytically for all quasinormal modes and for all black holes in the family we identified. Even more surprising is the fact that one can find a full analytic expression for the quasinormal modes of the lowest lying stable mode of black holes with $\alpha=2$ and $T \rightarrow 0$, with
$\alpha= \frac{J_{\phi}/N^2}{(Q/N^2)^2}$.
Such black holes are arbitrarily close to the supersymmetric black hole located at $T=0$.

\subsubsection*{The Emergence of a $1+1$ CFT and its Relation to the Conjectured Fermi Surface}
\begin{itemize}
\item It is interesting to compare the leading order of the large charge limit of the quasinormal modes we obtained to the quasinormal modes of a BTZ black hole. In the large charge limit, corresponding to $a \rightarrow 1$ on the gravity side, the dependence of $\lambda$ on $m_{\phi}$ and $\alpha$ vanishes to first order in $1-a$.
    Then, the spectrum is
    \begin{equation}
    \omega=m_{\phi}-i4\pi T \left(D(\alpha)+n\right).
    \end{equation}
    For comparison, the expressions obtained in \cite{Son:2002sd} for the quasinormal modes of a BTZ black hole are
    \begin{align}
    \omega_n^{(L)}= & k-i4\pi T_L \left(h_L+n\right) \\
    \omega_n^{(R)}= & -k-i4\pi T_R \left(h_R+n\right).
    \end{align}
    Indeed, the quasinormal modes we obtained look exactly like those of the left sector of the BTZ black hole, when replacing the fourier variable corresponding to momentum, $k$, with that corresponding to the angular momentum in our case, $m_{\phi}$.
\item Recall that, as shown in ref. \cite{Son:2002sd}, the quasinormal modes of perturbations around a black hole background correspond to the poles of the retarded Green's function on the CFT side. In the case of the BTZ black hole the dual CFT is a $1+1$ field theory. Thus, the structure of the quasinormal modes we found implies that our extremal black hole, in the large charge limit, matches the left sector of a $1+1$ CFT.
\item As discussed in the previous section, in this limit, as well as when $m_{\phi}=0$, there is always a zero eigenvalue
    for $\lambda$, and the spectrum corresponding to this eigenvalue is
    \begin{equation}
    \label{omega_discussion}
    \omega=m_{\phi}-i4\pi T \left(1+n\right).
    \end{equation}
    Thus, the lowest excitation is an operator with dimension $h_L=1$ in a $1+1$ CFT.
\end{itemize}

Given this last observation, we would like to examine the compatibility of the above results with our conjectured dual Fermi surface operators. We have suggested that this family of black holes corresponds to the $PSU(1,1|2)$ sector which is limited, in terms of derivatives, to $D_{1\dot{1}}$. This means that fields inside the sector are constrained to rotating with a specific chirality in the $\phi$ direction and effectively live in a chiral sector of a $1+1$ CFT, with the $x$ direction in the standard CFT replaced by $\phi$. Note that the perturbation we considered, the massless uncharged scalar $\Phi$, is inside the sector, as it has $m_{\psi}=0$, all charges equal to zero, and  $\omega=m_{\phi}$ (see relation \ref{psu_constraint} and \ref{angular momentum relation}). Thus, it is quite natural that we find poles corresponding to an operator living in the left sector of a chiral $1+1$ CFT, if one replaces the symmetry for translations along $x$ by symmetry for translations along $\phi$.

Furthermore, taking the large charge limit, we obtain the spectrum of a free fermion bilinear, $D(2)=1$, in such a theory.
Recall that $\Phi$ in $AdS_5 \times S^5$/${\cal N}=4$ SYM is dual to the lagrangian of the CFT.
When restricted to the $PSU(1,1|2)$ sector, only the fermion bilinears remain.
Hence, it is natural for the dilaton to excite a fermion bilinear in this $1+1$ chiral field theory.
The decay rate of these excitations is very small since it scales with the temperature $T$.
Therefore these excitations are quite long lived.

In the context of the Fermi surface operators, we are interested in values of $\alpha$ of the form
$\alpha=2+o(\frac{N}{Q})$ (see the discussion in the end of section \ref{Construction of Fermi surface subsubsection}).
Curiously, when $\alpha=2$, we find analytically the lowest mode for every value of $a$ and $m_{\phi}$.

Our results lend support to the conjecture that this degeneration of a charged rotating black hole
is described by a chiral $1+1$ CFT, which is a Fermi surface of the microscopic fermion. To verify this, the next step would be to compute the poles in fermionic Green's function, as well as those of additional, charged, scalar fields.
Not only should the results agree with those of a chiral $1+1$ CFT,
but the coupling of these bulk fields to the black hole are unambiguously prescribed by our model in Section \ref{black_hole_sec}.
We hope to return to this in future work.

At first glance, our results are reminiscent of the results obtained using Kerr/CFT \cite{Guica2009} and its
extensions to AdS, higher dimensions and gauged supergravities, e.g. \cite{Lu2009,Lu2009b}.
For a review, see \cite{Compere2012}.
In fact, Kerr/CFT has been applied to a broad family of extremal backgrounds in gauged supergravities \cite{Chow:2008dp}.
Our black hole is obtained as a limit of these backgrounds, but is not covered by these works as the extremal near horizon transformation used there becomes singular in our case.
More specifically, our black hole has zero entropy at extremality, and thus it is unclear how to apply the Kerr/CFT
duality in this case. As mentioned, the more recent work on EVH may be more appropriate \cite{Sheikh-Jabbari2011, Boer2011}.

It would be interesting to examine in more detail the near horizon
structure of this black hole when it becomes singular,
which we have not discussed in this work.
Understanding the singularity structure at $y=0$ and zero entropy degeneration at extremality
can help shed light on the dual low energy CFT.

Finally, The fact that we have obtained solutions in the form of Hypergeometric functions
suggests the presence of some $SL(2, \mathbb{R})$ symmetry in the near extremal limit, at least for the massless scalar in this background.
It may be useful to understand how the $SL(2, \mathbb{R})$
symmetry of the obtained solutions manifests itself in this scenario,
compared to previous results, such as \cite{Birkandan2011,Castro:2010fd, Krishnan:2010pv}, as well as
its possible connection to the $AdS_3$ factor present in the near horizon limit of the black hole away from the singularity.

\acknowledgments
We are grateful to J.~Simon for many suggestions and enlightening conversations,
and collaboration at early stages of this project.
We would also like thank S.~Minwalla, H.~Reall and O.~Aharony
for useful input at various stages of this project.
This work was supported in part by the Israel-U.S. Binational Science Foundation, by the Israel Science Foundation
(grant number 1665/10), and by the German-Israeli Foundation (GIF) for Scientific
Research and Development (grant number I-1-038-47.7/2009).
\newpage

\appendix
\section{Elimination of additional possible quantization conditions for \texorpdfstring{$\lambda$}{lambda}}
\label{Appendix_A}
First, examining the case $\rho_1 = \rho_2$ we discover that the equality in equation (\ref{zeroth}) is satisfied without further restrictions. This implies the relation $\rho_1 = \rho_2$ is exact, without any non analytic corrections. However, if we wish to consider this case we can't use the results achieved thus far since the linear transformations we used in equations (\ref{linear_1}) and (\ref{linear_2}) are no longer correct and there are other relations specific for this case. Repeating the process with the changed relations amounts to an equation including $ln(|s|)$ on the one side and a sum of Digamma functions on the other. Since Digamma functions can only have simple poles, this equation can never be satisfied and therefore we must discard this solution.

Thus, the remaining case is $\Re (\rho_1 - \rho_2) > 0$, for which we still haven't exhausted all the possibilities for equality in equations (\ref{radial_quant_4})-(\ref{radial_quant_3}).
Specifically, we need to check what happens for
$\rho_2-\rho_1=-n+O(s)$.
This corresponds to the special case we mentioned above, where $-d \in \mathds{N}$ for the Hypergeometric functions both in the far and near regions and so should be treated with more care.  In particular, before taking the limit $s\rightarrow 0$ in equations (\ref{zeroth near}), (\ref{zeroth far}), we should consider that, provided either $a$ or $b$ are not equal to some negative integer $-m$, satisfying $m<n$, the following holds
\begin{equation}
\lim_{d \rightarrow -k}\frac{F(a,b,d;z)}{\Gamma [d]}=\frac{(a)_{k+1}(b)_{k+1}}{(k+1)!}z^{n+1} {_2F_1}[a+k+1,b+k+1;k+2;z].
\end{equation}
We can now make use of $\rho_2-\rho_1+1=-k$, i.e $n=k+1$, along with the formula $\Gamma[-n+1]=-n\Gamma[-n]$ and divide equations (\ref{zeroth near}) and (\ref{zeroth far}) by $\Gamma[\rho_2-\rho_1]$. This allows us to take the limit $s\rightarrow 0$. For the zeroth order of equation (\ref{zeroth near}) we get
\small
\begin{align}
\label{zeroth near2}
F(t)=&\frac{\Gamma \left[\rho _1-\rho _2\right]\Gamma [\gamma ]}{\Gamma \left[\gamma +\rho _1\right]\Gamma \left[-\rho _2\right]}\left(\frac{t}{\sqrt{-s}}\right)^{\rho _2\text{   }}\frac{\left(-\rho _1\right)_n\left(-\gamma -\rho _1+1\right)_n}{(n-1)!}(-1)^{n+1}+ \nonumber \\ &
\frac{\Gamma [\gamma ]}{\Gamma \left[\gamma +\rho _2\right]\Gamma \left[-\rho _1\right]}\left(\frac{t}{\sqrt{-s}}\right)^{\rho_2},
\end{align}
\normalsize
whereas for the far region, equation (\ref{zeroth far}), we get
\small
\begin{align}
\label{zeroth far2}
F(t)&=\frac{-\Gamma \left[\rho _1-\rho _2\right]\Gamma [\delta ]}{\Gamma \left[\alpha +\rho _1\right]\Gamma \left[\beta +\rho _1\right]}\left(\sqrt{-s}t\right)^{\rho _1\text{   }}\frac{\left(\beta +\rho _2\right)_n\left(\alpha +\rho _2\right)_n}{(n-1)!}+\nonumber \\
&\left(\sqrt{-s}t\right)^{\rho _1}\frac{\Gamma [\delta ]}{\Gamma \left[\delta -\alpha -\rho _1\right]\Gamma \left[\delta -\beta -\rho _1\right]}.
\end{align}
\normalsize
The comparison between the two equations leads us to the conclusion that $\rho_2-\rho_1=-n$ doesn't constitute a solution since the matching of the two solutions is not possible.
Finally, we should recall that we have assumed neither $a$ nor $b$ of the problematic Hypergeometric functions in equations (\ref{zeroth near}) and (\ref{zeroth far}) are equal to a negative integer smaller in absolute value then $n$. We now consider the case where this assumption does not hold.

In this case, the Hypergeometric functions in question no longer diverge and we can use the previous expansion $F[a,b,d;s] \sim 1+O(s)$. However, comparing equations  (\ref{zeroth far}) and (\ref{zeroth near}) with (\ref{zeroth near2}) and (\ref{zeroth far2}), respectively, we see that if $a$ or $b$ fulfill this requirement only in one side of the equation, i.e only in the far or near region, we will again be unable to do the asymptotic matching since the powers of $t$ on both sides of the matching equation are different.
Now we can re-examine equation (\ref{zeroth}) in the case $a=-m$ or $b=-m$  and $n<m$ for both the far and the near region. For the far region we have
\begin{align}
\label{last_case}
a_{far}= & \beta+\rho_2 \nonumber \\
b_{far}=& \alpha+\rho_2
\end{align}
while for the near region
\begin{align}
a_{near}= & -\rho_1 \nonumber \\
_{near}=& -\rho_1-\gamma+1.
\end{align}
Now we can rewrite equation (\ref{zeroth}) as
\begin{equation}
\frac{\left(\Gamma \left[n\right]\right){}^2\Gamma \left[\gamma +\rho _2\right]\Gamma \left[a_{near}\right]\Gamma \left[a_{far}\right]\Gamma \left[b_{far}\right]}{\Gamma \left[\alpha +\rho _1\right]\Gamma \left[\beta +\rho _1\right]\left(\Gamma \left[-n\right]\right){}^2\Gamma \left[\gamma +\rho _1\right]\Gamma \left[-\rho _2\right]}=\left(-s\right)^{2n}
\end{equation}
We observe that, for this equality to be fulfilled, we need to have at least one more divergent gamma function in the denominator then in the numerator of the LHS since the RHS goes to zero. We will show this doesn't happen, assuming throughout our considerations the only divergent term in the denominator is $\left(\Gamma \left[-n\right]\right){}^2$ since the other cases were already treated.
Thus, we need to check only the cases where there is only one divergent gamma function in the numerator.
This can only happen when $a_{far}$, $b_{far}$ are equal to $-m_1,-m_2>-n$ and $b_{near}$ also fulfills this requirement but $a_{near}$ doesn't fulfill it.

We must assume then that (1)$\rho_1-\rho_2=n$ and (2) $-m_{\phi}-\rho_1>-n$. First, we observe the equality $\rho_2=-\rho_1-\gamma$ is true for the specific parameters of our equation. This implies that $\rho_2+1=-m$ and hence $\rho_2 \in \mathds{Z}$ and from (1) also $\rho_1 \in \mathds{Z}$.
Next, inserting $\gamma=2\mu+1$ into assumption (2), we find $-\rho_1>-n+2\mu$. However, we also need to have $\rho_1<-n$, since we showed it is an integer. It follows that $2\mu<0$, and since $\rho_1 \in \mathds{Z}$ we have $2\mu \in \mathds{Z}$.

Using (1) we shall now prove that the in this case $\Gamma[\gamma+\rho_2]$ is divergent.
\begin{equation*}
\rho_2+\gamma=1+n-\rho_1+2\mu < 1+2\mu < 1
\end{equation*}
since both $\gamma$ and $\rho_2$ are integers, obviously $\gamma+\rho_2$ is an integer, at most equal to zero. This ends our proof.

\section{Analytic expressions for the radial part of the eigenfunction}
\label{Appendix_B}
First we present the explicit expression for the radial part as a function of the separation constant $\lambda$ in the region near infinity , corresponding to $\sqrt{z_2} < r < 1$,
\begin{align}
& R(r)=(-1+r)^2 r^{\frac{1}{2} \left(-1+\sqrt{1-\frac{\alpha  \lambda }{\alpha +a (4+a \alpha )}}\right)} {_2F_1}\left[A,B,3,1-r\right] \\
& A=\frac{1}{2} \left(3+\sqrt{1-\frac{\alpha  \lambda }{\alpha +a (4+a \alpha )}}-\frac{\left(-1+a^2\right) \sqrt{1+a^2+\frac{4 a}{\alpha }} \alpha ^2 m_{\phi }}{(2+a \alpha ) (\alpha +a (4+a \alpha ))}\right) \\
& B= \frac{1}{2} \left(3+\sqrt{1-\frac{\alpha  \lambda }{\alpha +a (4+a \alpha )}}+\frac{\left(-1+a^2\right) \sqrt{1+a^2+\frac{4 a}{\alpha }} \alpha ^2 m_{\phi }}{(2+a \alpha ) (\alpha +a (4+a \alpha ))}\right).
\end{align}
where we have used our previous definition of
\begin{equation}
r=\frac{x^2}{x^2+(1+a)^2}.
\end{equation}
For the case of the first stable mode of a black hole with $\alpha=2$ we have obtained $\lambda$ analytically and can thus simplify the above expression to the form
\begin{align}
R(r)= &\frac{2 (1+a) r^{-\frac{m_{\phi }}{2}} \left((1+a) r^{\frac{m_{\phi }}{1+a}}+r^{\frac{a m_{\phi }}{1+a}} \left(-1-a+(-1+a) m_{\phi }\right)\right)}{(-1+a) \left(-1-a+(-1+a) m_{\phi }\right) m_{\phi }}+ \nonumber \\
& \frac{2 (1+a) r^{-1+\frac{(-1+a) m_{\phi }}{2 (1+a)}}}{1+a+(1-a) m_{\phi }}.
\end{align}

In the near horizon region , corresponding to $0 < r <\sqrt{z_2}$, we shall not present here the general expression but rather only that for the case of $\alpha=2$ where we obtained $\lambda$ analytically. It is given by
\footnotesize
\begin{align}
R(r) = & -\frac{2 (1+a)^{-2 n}\text{  }\Gamma \left[1-\frac{(-1+a) m_{\phi }}{1+a}\right] \Gamma \left[-1-n+\frac{(-1+a) m_{\phi }}{1+a}\right]{}^2\text{  }\text{Sin}\left[\frac{(-1+a) \pi  m_{\phi }}{1+a}\right] z_2^n}{\pi  \Gamma \left[-1-2 n+\frac{(-1+a) m_{\phi }}{1+a}\right]}(r-1)^2 r^{-1-n+\frac{(-1+a) m_{\phi }}{2 (1+a)}}\times \nonumber \\
& {_2F_1}\left[-n,-1-n+\frac{(-1+a) m_{\phi }}{1+a},-1-2 n+\frac{(-1+a) m_{\phi }}{1+a},\frac{z_1 r}{z_2}\right].
\end{align}
\normalsize
\bibliographystyle{JHEP}
\bibliography{blackholes}	

\end{document}